\documentclass[preprint,12pt]{elsarticle}

\usepackage{amssymb}

\usepackage{lineno}

\usepackage{cite} 
\usepackage{booktabs} 
\usepackage{microtype}
\usepackage[breaklinks]{hyperref}
\usepackage{geometry}
\usepackage{pdflscape}
\usepackage{adjustbox}
\usepackage{placeins}
\usepackage{amsmath}
\usepackage{array}
\usepackage{subfig}
\usepackage{natbib} 

\journal{arXiv}

\begin{document}

\begin{frontmatter}



\title{FUELVISION: A Multimodal Data Fusion and Multimodel Ensemble Algorithm for Wildfire Fuels Mapping}



\author[inst1]{Riyaaz Uddien Shaik}

\affiliation[inst1]{organization={Civil \& Env. Engineering Dept.},
            addressline={University of California Los Angeles}, 
            city={Los Angeles},
            postcode={90095}, 
            state={CA},
            country={USA}}

\author[inst2]{Mohamad Alipour}
\author[inst3]{Eric Rowell}
\author[inst4]{Bharathan Balaji}
\author[inst5]{Adam Watts}

\author[inst1]{Ertugrul Taciroglu\texorpdfstring{\corref{cor1}}{}}
\ead{etacir@ucla.edu}
\cortext[cor1]{Corresponding author}

\affiliation[inst2]{organization={Department of Civil and Environmental Engineering},
            addressline={University of Illinois at Urbana-Champaign}, 
            city={Urbana},
            postcode={61801}, 
            state={IL},
            country={USA}}

\affiliation[inst3]{organization={Division of Atmospheric Sciences},
            addressline={Desert Research Institute}, 
            city={Reno},
            postcode={89512}, 
            state={NV},
            country={USA}}

\affiliation[inst4]{organization={Amazon},
            city={Seattle},
            postcode={98109}, 
            state={WA},
            country={USA}}

\affiliation[inst5]{organization={Pacific Wildland Fire Sciences Laboratory},
            addressline={United States Forest Service}, 
            city={Wenatchee},
            postcode={98801}, 
            state={WA},
            country={USA}}
            
\begin{abstract}
\noindent
Accurate assessment of fuel conditions is a prerequisite for fire ignition and behavior prediction, and risk management. The method proposed herein leverages diverse data sources---including Landsat-8 optical imagery, Sentinel-1 (C-band) Synthetic Aperture Radar (SAR) imagery, PALSAR (L-band) SAR imagery, and terrain features---to capture comprehensive information about fuel types and distributions. An ensemble model was trained to predict landscape-scale fuels---such as the ``Scott and Burgan 40'' or time-lag 1hr, 10hr, and 100hr fuels---using the as-received Forest Inventory and Analysis (FIA) field survey plot data obtained from the USDA Forest Service. However, this basic approach yielded relatively poor results due to the inadequate amount of training data. Pseudo-labeled and fully synthetic datasets were developed using generative AI approaches to address the limitations of ground truth data availability. These synthetic datasets were used for augmenting the FIA data from California to enhance the robustness and coverage of model training. The use of an ensemble of methods---including deep learning neural networks, decision trees, and gradient boosting---offered a fuel mapping accuracy of nearly 80\%. Through extensive experimentation and evaluation, the effectiveness of the proposed approach was validated for regions of the 2021 Dixie and Caldor fires. Comparative analyses against high-resolution data from the National Agriculture Imagery Program (NAIP) and timber harvest maps affirmed the robustness and reliability of the proposed approach, which is capable of near-real-time fuel mapping.

\end{abstract}



\begin{keyword}
Wildfires \sep Fuel Mapping \sep Artificial Intelligence \sep Ensemble Model \sep Synthetic Data Generation

\end{keyword}

\end{frontmatter}



\section{Introduction}
\noindent
Recent studies indicate an unparalleled rise in the magnitude, severity, and impact of wildfire occurrences [\citenum{burke2021changing, iglesias2022us}]. In 2018, California witnessed the deadliest fire in its history, the Camp Fire, resulting in the loss of 85 lives, and the destruction of nearly 14,000 homes and over 500 commercial structures [\citenum{iglesias2022us}]. With the exacerbation of these incidents due to climate change, the United Nations Environment Program projects a further global increase of approximately 30\% by 2050 and 50\% by the end of the century [\citenum{sullivan2022spreading}]. Despite advances in fire science, both technologically and theoretically, wildfires persist as a significant and escalating threat to communities, infrastructure, and the environment. The unprecedented scale and complexity of this issue necessitate interdisciplinary and data-informed research on wildfire risk management, encompassing assessment, mitigation, and response strategies.

Efficient wildfire risk management relies on accurate simulations of wildfire spread, as these simulations can significantly enhance the effectiveness of pre-event mitigation, evacuation, rescue, and fire suppression efforts [\citenum{kalabokidis2016aegis, sakellariou2017review}]. An essential component of wildfire simulations is obtaining reliable estimates of the fuels that contribute to the spread of wildfires. Fuels are typically classified into three categories: ground fuels (including litter, duff, and coarse woody debris), surface fuels (such as grass, forb, shrubs, and large logs), and canopy fuels (consisting of trees and snags) [\citenum{keane2015wildland}]. While surface fuels play a primary role in initiating and propagating forest fires, in this research, we are considering mapping the '40 Scott and Burgan' standard fuel models [\citenum{scott2005standard}], which were the primary input for point-based and spread simulations and were derived from the 'Anderson 13' categorization standard fuel models [\citenum{anderson1981aids}].

Methods for characterizing surface fuels have been developed in general, failing to capture the full range of temporal variability and spatial non-conformity inherent to the surface fuel beds [\citenum{keane2015wildland}]. Consequently, the input data for modern fire behavior models contain uncertainties in describing the dynamic processes that traditional fuel inventories miss [\citenum{rowell2016using}]. A review of the current state of surface fuel mapping research reveals that past efforts have predominantly focused on site-specific semi-manual expert systems or traditional machine learning methods (such as decision trees and random forests) at regional scales (approximately 30km x 30km). However, these systems have limited capabilities in harnessing big data analytics, which could be leveraged to extract knowledge from spatial and spectral consistencies and ensure consistent vegetation and fuel assessment across a given landscape. Consequently, such systems experience a decrease in prediction accuracy when attempting to generalize their results to larger problem domains such as state or nation-wide.

LANDFIRE [\citenum{rollins2009landfire}] provides comprehensive and consistent geospatial fuel map products at the national level (CONUS), which were initially mapped in 2016 by integrating remote sensing, machine learning, expert-driven rule sets, and quality control. These products are updated every two years to incorporate data on disturbances such as deforestation and fires. While these products have provided a valuable foundation for fire spread simulation efforts, there is a need for large-scale modeling techniques that can generate geo-referenced fuel mapping in near-real-time (within the same season) and without relying solely on experience-driven rule-sets and localized vegetation models [\citenum{keane2011use}]. Implementing such models enhance the frequency and reduce the time delay of fuel data, which currently takes several years (~2 years) to update. Additionally, new techniques could facilitate a comprehensive and systematic accuracy assessment using independent validation datasets.

Taking into account all above-mentioned limitations, this paper presents an AI-based framework that incorporates multiple modalities of data, including multi-spectral satellite imagery, C-band SAR data, L-band SAR data, and terrain data. The framework relies on a combination of ensemble and stacked machine learning models, which are trained using state-wide georeferenced labeled data. The trained ensemble model can be utilized to identify on-demand near-real-time fuels (within the same season).

\subsection{Background}
\noindent Fuel mapping research is underway worldwide, but the primary focus has been on developing fuel map products. Efforts to develop algorithms that can generate real-time, on-demand fuel maps have been sparse. Notable examples include the study by Pickell et al. [\citenum{pickell2020fuelnet}], who have developed FuelNet, which is an artificial neural network-based algorithm for updating existing fuel maps in Canada (ca. 2016). They utilized remotely sensed satellite imagery to create updated fuel maps and achieved an overall accuracy of approximately 63.1\%. Shaik et al. [\citenum{shaik2022automatic}] have developed a semi-supervised algorithm using PRISMA hyperspectral imagery to map fuel types across Europe, achieving an overall accuracy of 87\%. However, this algorithm does not directly map fuels; instead, it maps vegetation types and correlates them with the 'Anderson 13' categorization of standard fuel models [\citenum{anderson1981aids}]. Furthermore, its effectiveness is limited by the availability of PRISMA hyperspectral imagery.

Before undertaking the present effort, a preliminary study was carried out to develop a deep learning framework capable of processing multimodal data for large-scale surface fuel mapping [\citenum{alipour2023multimodal}]. A multi-layer neural network was employed to incorporate both spectral and biophysical data, while a convolutional neural network backbone was utilized to extract visual features from high-resolution imagery. A Monte Carlo dropout mechanism was also developed to generate a stochastic ensemble of models, which effectively captured classification uncertainties while enhancing prediction performance. To demonstrate the system's efficacy, fuel pseudo-labels were generated by randomly sampling the LANDFIRE fuel map across California as a proof-of-concept. This prior effort paved the way for the current study by demonstrating the feasibility of fuel mapping for the entire state of California and indicated that the method can produce an operationalizable fuel mapping tool by incorporating field data derived from in-situ surveys, such as the Forest Inventory and Analysis (FIA) database [\citenum{alipour2023multimodal}].

\subsection{Research Significance}
\noindent In this work, emerging machine learning techniques are leveraged to develop an algorithm for real-time, on-demand fuel maps for any selected domain. To that end, a data fusion scheme is devised to integrate optical, synthetic aperture radar (SAR), and terrain data and to identify fuels using a single end-to-end model for the state of California. To create a training dataset, fuel labels from FIA plots [\citenum{FIAv9}]  are coupled with multimodal input data sourced from various data repositories and geospatial data products, including multispectral satellite data (time-series NDVI), C-band SAR data (VV and VH), L-band SAR data (HH and HV), SAR-based spectral indices and topography and terrain data (from the U.S. Geological Survey (USGS) Digitial Elevation model). The dataset size has been improved by pseudo-labeling and synthetic data augmentation. We trained the machine learning model using the created dataset and named it as the FuelVision. The proposed approach presents the following technical contributions and benefits with respect to the existing literature:
1. Creating fuel identification models that are applicable to any selected domain with a spatial resolution of 30m while integrating multispectral, two types of SAR and terrain information and providing a measure of model uncertainty.
2. Creating a fuel identification model that can generate on-demand near real-time fuel maps.

The methodology and results present a detailed analysis of the effect of the individual components of the FuelVision model, the multi-model ensemble approach, pseudo-labeling, and synthetic data augmentation on the training dataset. Pseudo-labels and synthetic data augmentation demonstrate proof-of-concept, and examine the feasibility of fuel identification models in California.

\section{Study Area and Data Used}
\noindent This section provides an overview of the geographical scope and data sources utilized in this study. The analysis conducted in this research focuses on a specific region, referred to as the Region of Analysis. Since we obtained fuel labels from FIA plots, this section further explores the valuable insights gained from the analysis of FIA plots within the study area, allowing for a comprehensive understanding of the plots and their distribution across the region of interest. Additionally, the section delves into the utilization of remote sensing data and spectral information extracted from the remote sensing data.

\subsection{Region of Analysis}
\begin{figure}[htbp]
  \centering
  \includegraphics[width=0.9\textwidth]{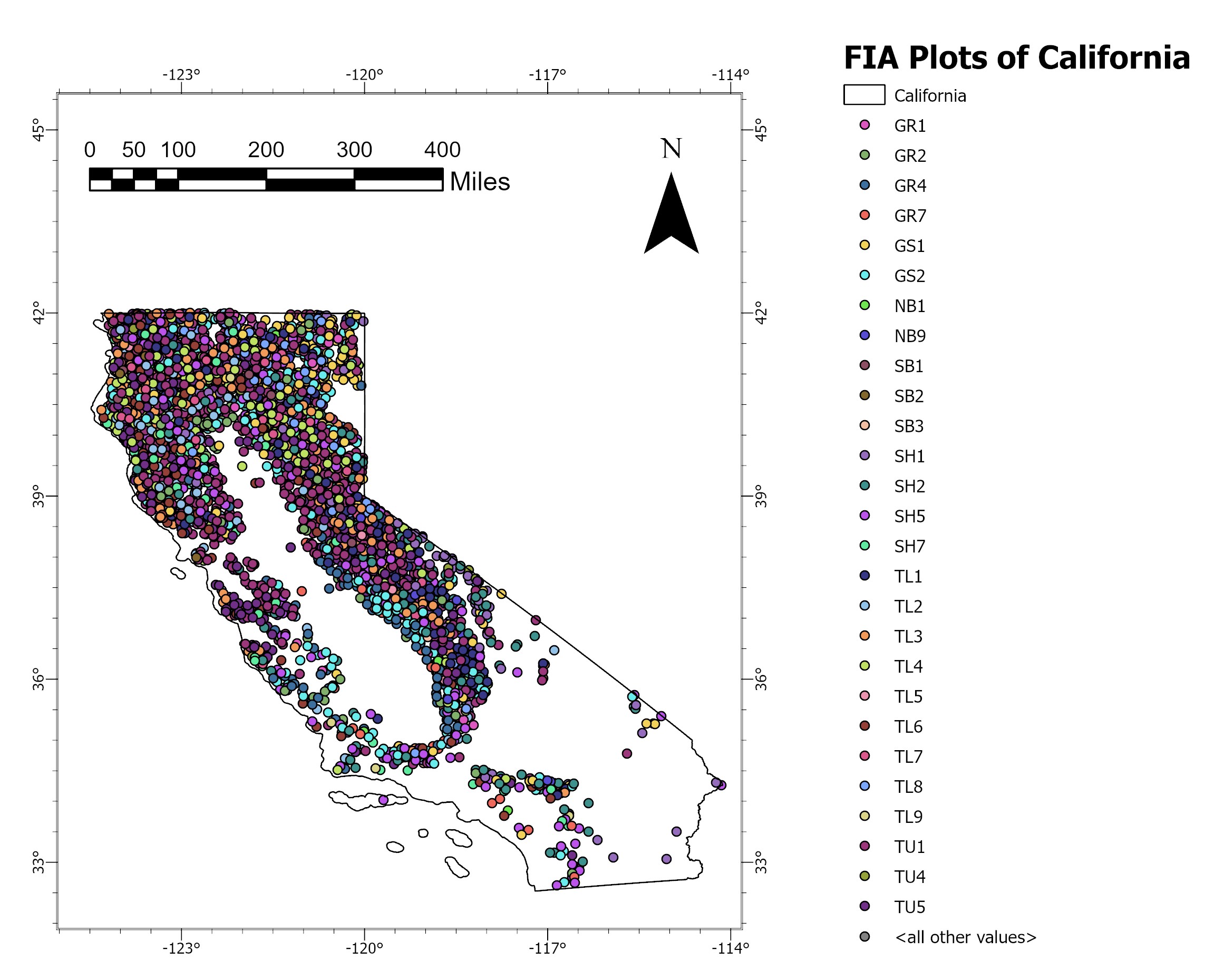} 
  \caption{FIA Plots with assigned fuel models in the study area of California (locations here are approximate to maintain FIA spatial confidentiality).}
  \label{fig:ROI}
\end{figure}

\noindent The region selected for data sampling and modeling in the present study is California. Figure \ref{fig:ROI} and Table \ref{tab:scott_burgan} depict the distribution of fuel labels we had for California and fuel models explanation, respectively. California offers a diverse range of elevations, ranging from -86 m in Death Valley to 4421 m at the summit of Mt. Whitney. Approximately 40\% of California's total area, which amounts to around 13.35 million hectares, is covered by forests [\citenum{bytnerowicz1996nitrogen}]. Moreover, the state boasts a rich assortment of ecosystems, including alpine, montane, and subalpine forests, coastal forests, mixed conifer-deciduous forests, chaparral, pinyon-juniper woodlands, and desert scrub. Due to this wide variety of forest ecotypes and the fact that California is frequently affected by wildfires, it serves as an ideal case study for this research.

\FloatBarrier

\begin{table}[htbp]
    \centering
    \caption {Fuel type description based on the Scott and Burgan fuel models adapted from [\citenum{Scott2005}]}
    \label{tab:scott_burgan}
    \begin{tabular}{p{0.9cm} p{13cm}}
        \toprule
        \textbf{FM} & \textbf{Fuel Description}\\
        \midrule
        GR1 & Grass: The grass is short, patchy, and possibly heavily grazed. The spread rate is moderate; flame length is low.\\
        GR2 & Grass: Moderately coarse continuous grass with an average depth of about 1 foot. Spread rate high; flame length  moderate \\
        GR4 &  Grass: Moderately coarse continuous grass, average depth is about 2 feet. Spread rate very high; flame length high.\\
        GR7 &  Grass: Moderately coarse continuous grass, average depth is about 3 feet. Spread rate very high; flame length very high.\\
        GS1 & Grass-Shrub: are about 1 foot high with a low grass load. Spread rate moderate; flame length low\\
        GS2 & Grass-Shrub: Shrubs are 1 to 3 feet high, with moderate grass load. Spread rate high; flame length moderate.\\
        SB1 & Slash-Blowdown: Fine fuel load is 10 to 20 tons/acre, weighted toward fuels 1 to 3 inches diameter class, depth is less than 1 foot. Spread rate moderate; flame length low.\\
        SB2 & Slash-Blowdown: Fine fuel load is 7 to 12 tons/acre, evenly distributed across 0 to 0.25, 0.25 to 1, and 1 to 3-inch diameter classes, depth is about 1 foot. Spread rate moderate; flame length moderate.\\
        SH1 & Shrub: Low shrub fuel load, fuel bed depth of about 1 foot; some grass may be present. Spread rate very low; flame length  very low.\\
        SH2 & Shrub: Moderate fuel load (higher than SH1), depth is about 1 foot, no grass fuel present. The spread rate is low; flame length is low. \\
        SH5 & Shrub: Heavy shrub load, depth 4 to 6 feet. The spread rate is very high; flame length is very high.\\
        SH7 & Shrub: Very heavy shrub load, depth 4 to 6 feet. The spread rate is lower than SH5, but the flame length is similar. The spread rate is high; flame length is very high.\\
        TL1 & Timber Litter: Light to moderate load, fuels 1 to 2 inches deep. The spread rate is very low; flame length is very low. \\
        TL2 & Timber Litter: Low load, compact. The spread rate is very low; flame length is very low\\
        TL3 & Timber Litter: Moderate load conifer litter. The spread rate is very low; flame length is low.\\
        
        \bottomrule
    \end{tabular}
\end{table}
\FloatBarrier

\begin{table}[htbp]
    \centering
    \captionsetup{} 
    \label{tab:scott_burgan2}
    \begin{tabular}{p{0.9cm} p{13cm}}
        \toprule
        \textbf{FM} & \textbf{Fuel Description}\\
        \midrule       
        TL4 & Timber Litter: Moderate load, including small-diameter downed logs. The spread rate is low; flame length is low.\\
        TL5 & Timber Litter: High load conifer litter; light slash or mortality fuel. The spread rate is low; flame length is low.\\
        TL6 & Timber Litter: Moderate load, less compact. The spread rate is moderate; flame length is low.\\
        TL7 & Timber Litter: Heavy load, including larger-diameter downed logs. The spread rate is low; flame length is low.\\
        TL8 & Timber Litter: Moderate load and compactness may include a small amount of herbaceous load. The spread rate is moderate; flame length is low.\\
        TL9 & Timber Litter: Very high load, fluffy. Spread rate moderate; flame length moderate.\\
        TU1 & Timber Understory: Fuel bed is low-load grass and/or shrub with litter. The spread rate is low; flame length is low.\\
        TU4 & Timber Understory: Fuelbed is short conifer trees with grass or moss understory. Spread rate moderate; flame length moderate.\\
        TU5 & Timber Understory: The fuel bed is a high-load conifer litter with shrub understory. The spread rate is moderate; flame length is moderate.\\
        \bottomrule
    \end{tabular}
\end{table}
\FloatBarrier

\subsection{FIA and its plots analysis}
\noindent The USDA Forest Service's Forest Inventory and Analysis (FIA) program conducts a comprehensive national inventory of forests in the United States. FIA stands as the sole program responsible for collecting, publishing, and analyzing data concerning forest land ownership across the nation. The spatial sampling intensity roughly equates to one plot per 6,000 acres. Each plot consists of four fixed-radius subplots, each measuring 24 feet, arranged in a clustered configuration. On each subplot, information about the stand and site is collected, including metrics like standing live/dead tree height/diameter and physiographic class/ownership. A subset of FIA's permanent inventory plots is sampled annually to assess indicators of forest health, including soils, understory vegetation, and down woody materials (DWM). The DWM indicator offers estimates regarding down and deceased woody materials within forest ecosystems. These DWM estimates play a crucial role in evaluating forest ecosystem attributes like fuel loadings, carbon stocks, and structural diversity. As defined by the FIA program, DWM encompasses both fine and coarse woody debris, slash piles, duff, litter, and cover and height of shrubs/herbs.

\begin{figure}[htbp]
  \centering
  \includegraphics[width=0.7\textwidth]{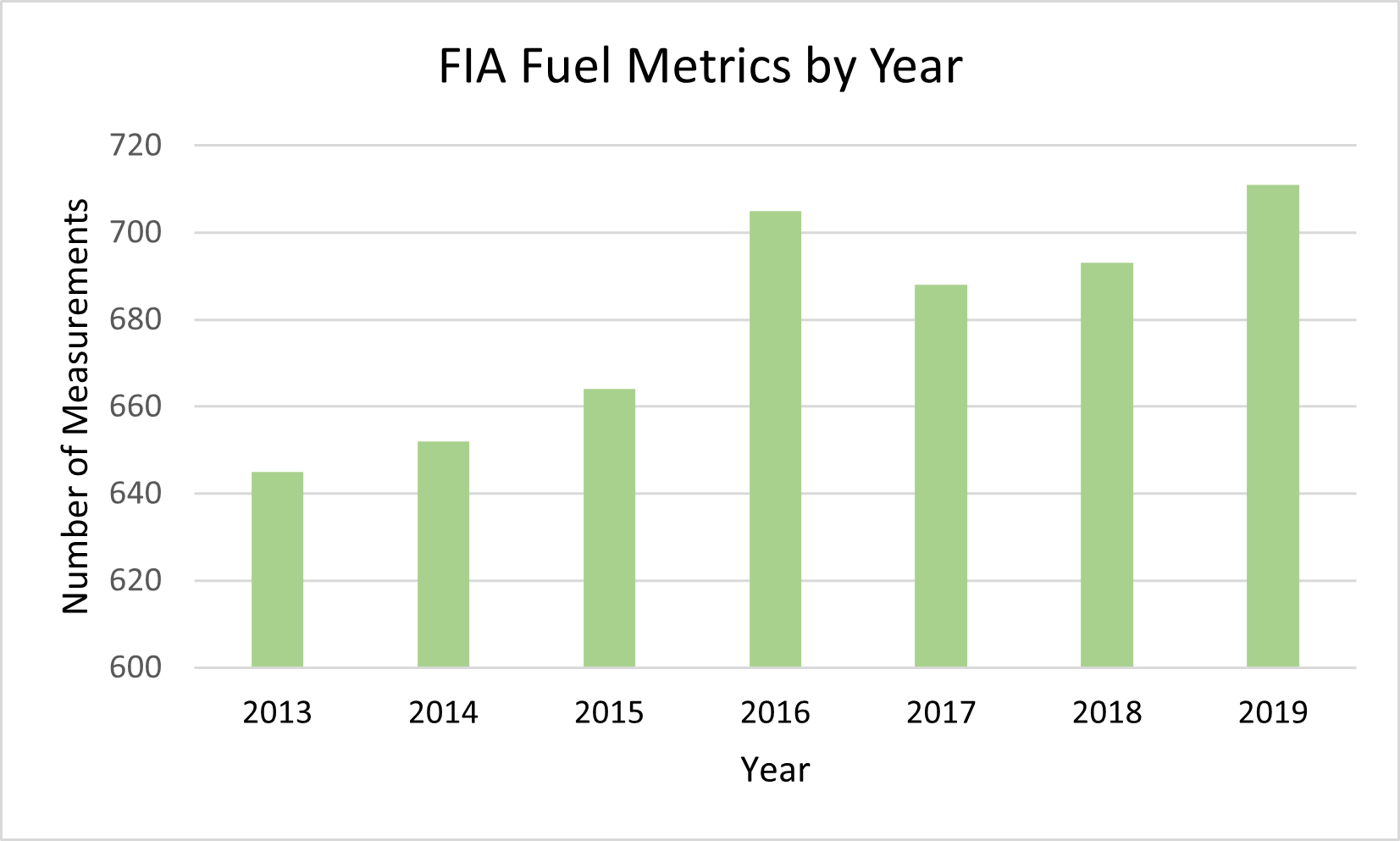} 
  \caption{Number of FIA plots with fuel data assignment per year}
  \label{fig:Annual_Fuel_Metrics}
\end{figure}

The bar graph presented in Figure \ref{fig:Annual_Fuel_Metrics} displays the number of FIA plots with fuel type information collected from the years 2013 to 2019 [\citenum{FIAv9}]. The purpose of this graph is to provide an overview of the temporal distribution and sampling effort of FIA plots during the specified time period. However, since satellite imagery is only available from 2015 onwards, only FIA plots from 2015 to 2019, totaling 3461, were utilized. 

\FloatBarrier

\begin{figure}[htbp]
  \centering
  \includegraphics[width=\textwidth]{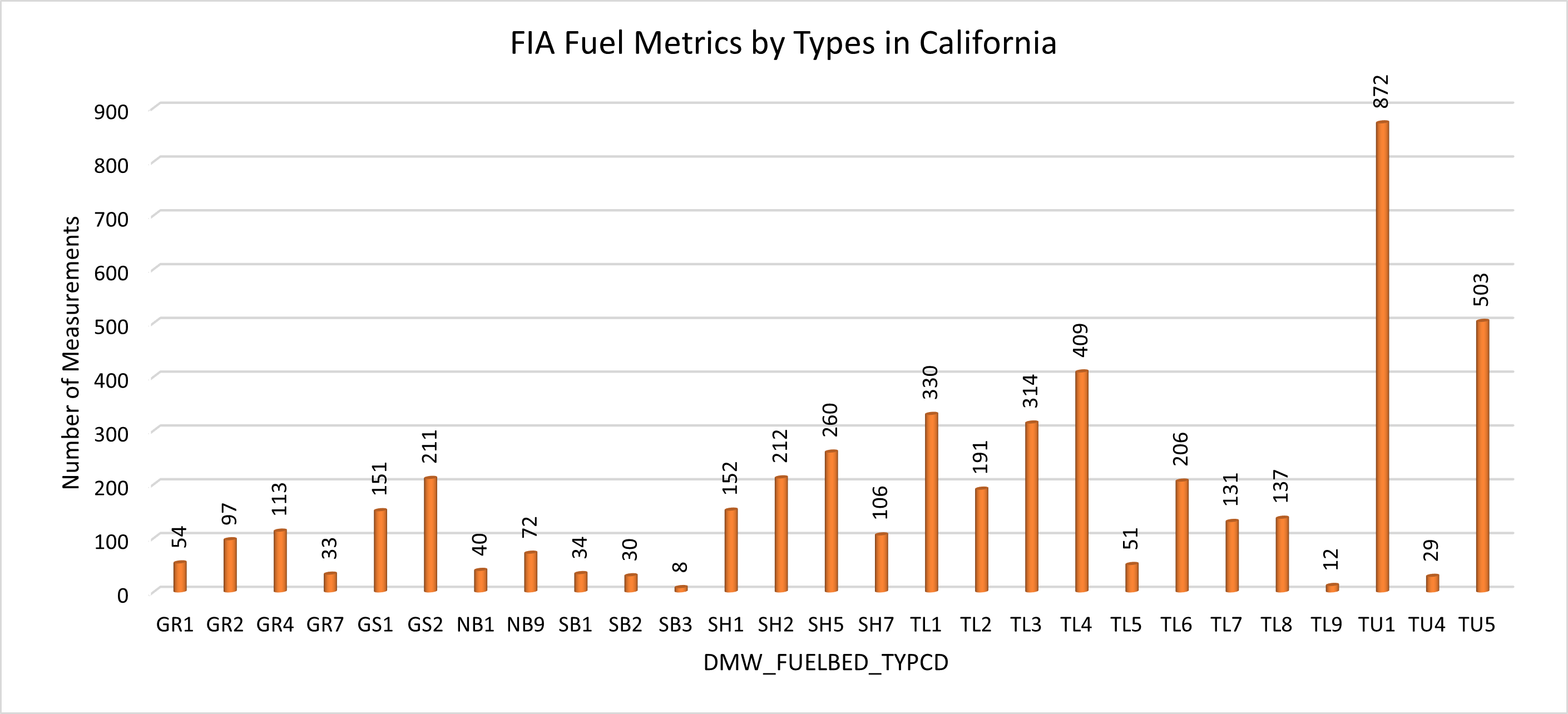} 
  \caption{Number of Fuel Models assignment in FIA plots for California}
  \label{fig:FIAFuelModels}
\end{figure}

Figure \ref{fig:FIAFuelModels} illustrates the distribution of different fuel model types in FIA plots. Among the fuel models, the category SB3 represents the minimum count, while TU1 exhibits the maximum count. One of the reasons for this disparity could be the sole consideration of forested lands (i.e., those with at least 10 percent canopy cover by live tallied trees of any size) for inventory analysis under FIA. This disparity indicates that the ground truth data we possess is imbalanced, highlighting the uneven representation of fuel labels in the dataset. 

\FloatBarrier

\subsection{Remote Sensing Data}
\noindent The input variables for the machine learning model architecture are derived from a combination of four distinct open-source remote sensing datasets accessible in the Google Earth Engine (GEE). These datasets, namely Landsat-8 optical imagery, Sentinel-1 Synthetic Aperture Radar (SAR) data, Phased Array type L-band Synthetic Aperture Radar (PALSAR) data, and Shuttle Radar Topography Mission (SRTM) elevation and slope data, are merged at various spatial and temporal resolutions to generate individual arrays, each approximating a spatial resolution of 30 x 30 meters.

\begin{table}[htbp]
    \centering
    \caption{Remote sensing datasets used in this study }
    \label{tab:input_data}
    \begin{tabular}{p{2.5cm} p{5cm} p{5cm}}
        \toprule
        \textbf{Data Type} & \textbf{Spectral Data} & \textbf{Spatial Resolution (m)} \\
        \midrule
        Landsat-8 & Two-years time-series NDVI & 30 \\
        Sentinel-1 & VV and VH Polarization & 10 \\
        PALSAR & HH and HV Polarization & 25 \\
        SRTM & Elevation and Slope & 30\\
        \bottomrule
    \end{tabular}
\end{table}
\FloatBarrier

The table displays various types of imagery, including Landsat-8, Sentinel-1, PALSAR, and SRTM, along with their corresponding spectral data and resolution. The worldwide utilization of these data for land use land cover classification (LULC) applications can be attributed to three key characteristics: free availability [\citenum{wang2017fusion}], interoperability [\citenum{wulder2015virtual}], and the ability to monitor expansive regions [\citenum{piedelobo2019scalable}]. These factors contribute to their high demand as crucial inputs for LULC analysis on a global scale.

Landsat-8 imagery provides multispectral data with a 30-meter resolution, enabling analysis across various spectral bands ranging from 442nm to 1373nm. With a temporal resolution of 16 days and a radiometric resolution of 16 bits, it offers valuable insights [\citenum{ed2020recent}]. Each scene covers an area of 185 × 180 km and is captured from sun-synchronous orbits, making it comparable in terms of spectral, spatial, and angular characteristics to the referenced source [\citenum{ed2020recent}]. A two-year seasonal time series of Normalized Difference Vegetation Index (NDVI) was generated for the period between 2015 and 2019. This time series captures the variations in NDVI throughout different seasons over the course of two years.

The Sentinel-1 (S1) SAR GRD collection available in GEE was utilized because it comprises radiometrically calibrated and terrain-corrected scenes [\citenum{mullissa2021sentinel}]. Initially, an image collection encompassing all accessible S1 scenes from 2015 to 2019 was gathered in the Interferometric Wide Swath mode, encompassing vertical-vertical (VV) and vertical-horizontal (VH) polarizations at a resolution of 30 m from cross orbits [\citenum{ban2020near}]. Subsequently, biannual mosaics were generated using medians. Six effective polarimetric features, namely SR-1 [\citenum{sayedain2020assessing}], SR-2 [\citenum{sayedain2020assessing}], Power Ratio (PR) [\citenum{sayedain2020assessing}], Total Scattering Power (SPAN) [\citenum{sayedain2020assessing}], Difference Intensity (DI) [\citenum{sayedain2020assessing}], Radar Vegetation Index (RVI) [\citenum{periasamy2018significance}]  and C-band Normalized Polarized Difference Index (C-NPDI) [\citenum{afifi2019statistical}], were identified as appropriate features for consideration as shown in table 2. These features were selected due to their relevance and potential impact on vegetation classification.

The Japan Aerospace Exploration Agency (JAXA) provides the ALOS/PALSAR yearly mosaic at a resolution of 25 m [\citenum{chen2021forest}]. This mosaic is created by merging SAR images obtained from either PALSAR-1 or PALSAR-2, available for each year [\citenum{xu2022improving}]. This SAR imagery underwent orthorectification and slope correction using the 90m SRTM Digital Elevation Model. To address intensity differences caused by variations in surface moisture conditions, a destriping process was employed. This process equalized the intensities between neighboring strips, which often arise from seasonal and daily fluctuations in surface moisture [\citenum{huang2019constructing}]. Initially, the data were in digital number (DN) format and were later converted to gamma-naught ($\gamma^0$) values using Eq.(1) [\citenum{huang2019constructing}] within the GEE platform.

\begin{center}
\begin{equation}
\gamma^0 = 10 \times \log_{10} (DN^2) - 83
\end{equation}
\end{center}

\begin{table}[htbp]
    \centering
    \caption{Spectral Indices Used as Training Features}
    \label{tab:spectral_indices}
    \renewcommand{\arraystretch}{1.5} 
    \begin{tabular}{p{1.5cm} p{3.5cm} p{1.2cm} p{4.5cm} p{2cm}}
        \toprule
        \textbf{Index} & \textbf{Formula} & \textbf{Source} & \textbf{Application} & \textbf{Ref} \\
        \midrule
         NDVI & $\frac{NIR-R}{NIR+R}$ & L8 & Vegetation dynamics over time & [\citenum{chang2019chimera}]] \\
         SR-1 & $\frac{VH}{VV}$ & S1 & Vegetation separability & [\citenum{vhvv}] \\
         SR-2 & $\frac{VV}{VH}$ & S1 & Surface vegetation growth & [\citenum{koley2022sentinel}]\\
         PR & $\frac{\lvert VV\rvert^2_{\text{dB}}}{\lvert VH\rvert^2_{\text{dB}}}$ & S1 & land cover classification & [\citenum{sayedain2020assessing}]\\

         SPAN & $\frac{1}{2}(|VV|^2+|VH|^2)$  & S1 & land cover classification & [\citenum{sayedain2020assessing}] \\
         DI & $\frac{1}{2}(|VV|^2 - |VH|^2)$ & S1 & land cover classification & [\citenum{sayedain2020assessing}] \\
         RVI & $\frac{4 x VH}{VV + VH}$ & S1 & Vegetation and bare soil separability & [\citenum{periasamy2018significance}] \\
         C-NPDI & $\frac{VV-VH}{VV+VH}$ & S1 & Vegetation optical depth & [\citenum{li2022exploring}]\\
         L-NPDI & $\frac{VV-VH}{VV+VH}$ & PL & Vegetation optical depth & [\citenum{li2022exploring}]\\
         ESPRIT & $\frac{HH+HV}{2}$ & PL & Sensitive to vegetation height & [\citenum{liu2021estimation}]\\
         L-DIFF & ${HH-HV}$ & PL & Sensitive to vegetation height & [\citenum{chen2021forest}]\\
         C-Ratio & $\frac{HH}{HV}$ & PL & Sensitive to vegetation height & [\citenum{chen2021forest}]\\
        \bottomrule
    \end{tabular}
\end{table}
\FloatBarrier

Four effective polarimetric features based on PALSAR, namely L-band Normalized Polarized Difference Index (L-NPDI) [\citenum{afifi2019statistical}], Estimation of Signal Parameters via Rotational Invariance Techniques (ESPRIT) [\citenum{liu2021estimation}], L-band Difference (L-DIFF) [\citenum{chen2021forest}], and C-band Ratio (C-Ratio) [\citenum{chen2021forest}], were identified as suitable candidates for analysis, as depicted in Table 2.

The Shuttle Radar Topography Mission (SRTM) originally produced the digital elevation dataset to offer reliable, high-quality elevation data on a near-global scale with a spatial resolution of 30 m [\citenum{uuemaa2020vertical}]. GEE has processed the SRTM digital elevation data to fill in data gaps and enhance its usability [\citenum{f3fe31dd7b2d4983abfc45d2083a4493}]. By utilizing this elevation data, the slope is calculated on GEE and added as a predictor variable.

\section{Methodology}
\noindent In this section, a flowchart is presented in figure \ref{fig:FuelVision_v2} to illustrate the step-by-step procedures of the proposed fuel mapping method. The flowchart consists of five main stages. First, the input data are collected and filtered to ensure their relevance and quality. Next, a pre-processing step is carried out to clean and standardize the data, eliminating any inconsistencies or noise that could potentially affect the analysis. In this step, various indices are calculated, as presented in Table 2, and a cross-orbit data merging step is performed for the Sentinel-1 data, as described in detail earlier. Next, data augmentation and dataset preparation techniques are employed to expand the dataset, increasing its diversity and improving the model's generalization capabilities. Thereafter, the ensemble model is trained using the augmented dataset, leveraging its automated machine-learning capabilities to optimize model performance. The trained model is then evaluated to assess its accuracy and effectiveness in capturing the underlying patterns in the data. Finally, the fuel map generation stage utilizes the trained model to generate comprehensive fuel maps—--herein, for the Dixie and Caldor fire impact regions, as examples.

\newgeometry{margin=1in}
\begin{landscape}
\begin{figure}[htbp]
  \centering
    \begin{adjustbox}{max width=\linewidth}
      \includegraphics[width=25cm, height=15cm]{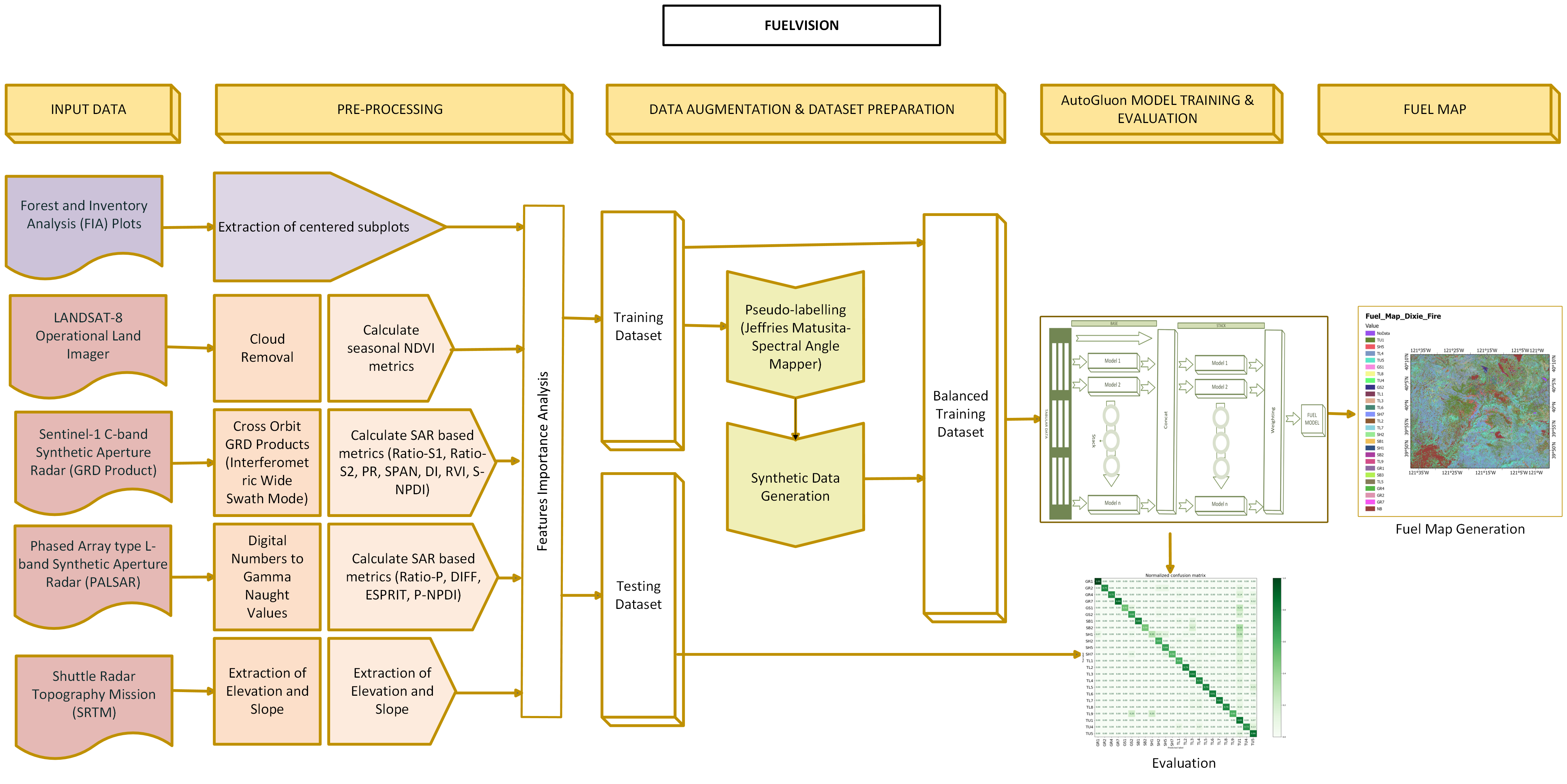} 
    \end{adjustbox}
    \caption{FuelVision Framework}
  \label{fig:FuelVision_v2}
\end{figure}
\FloatBarrier

\end{landscape}
\restoregeometry

\subsection{Label Propagation}
\begin{figure}[htbp]
  \centering
  \includegraphics[width=10cm, height=7.5cm]{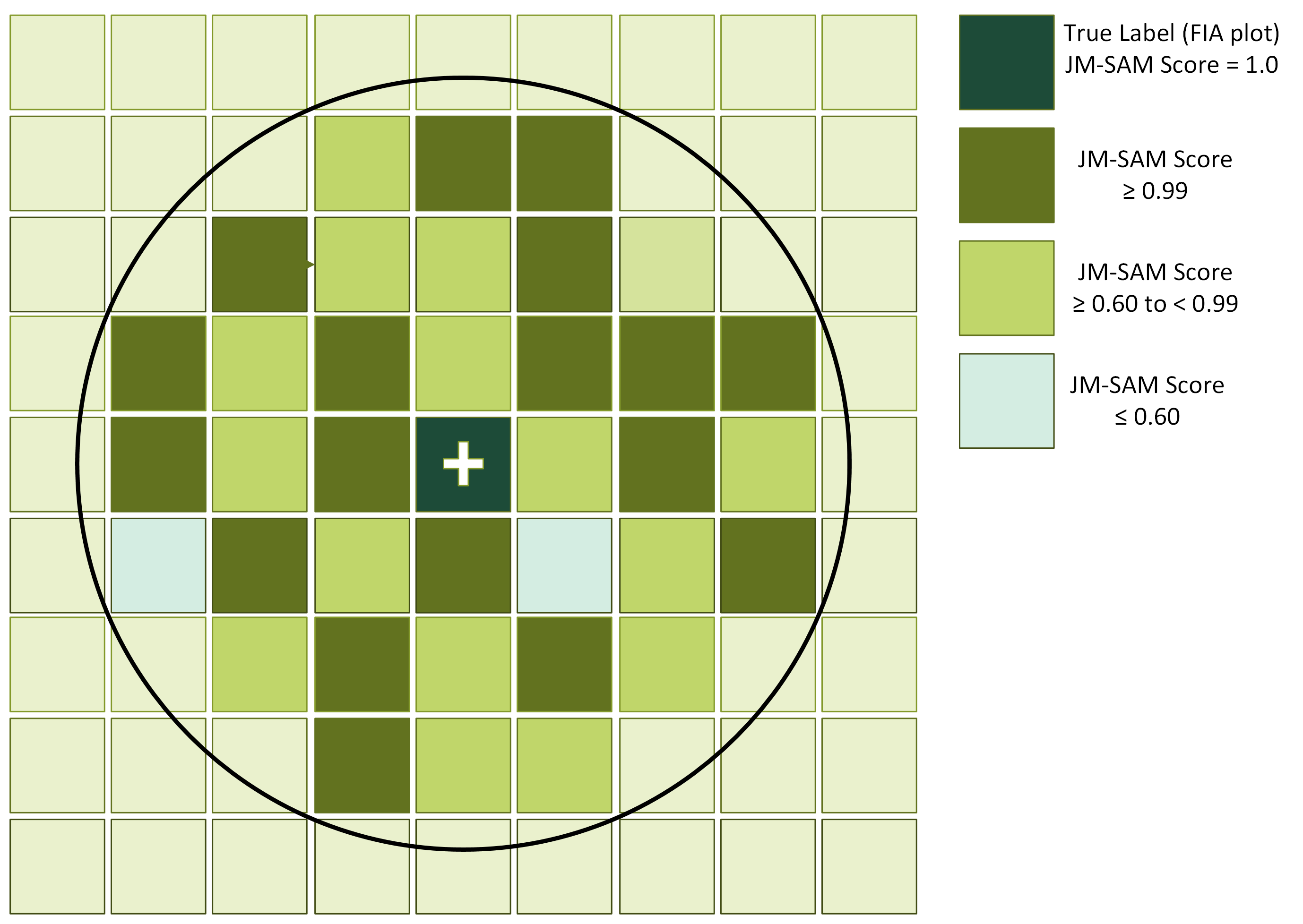} 
  \caption{Pseudo-Labelling using Jeffries-Matusita-Spectral Angle Mapping Scores}
  \label{fig:Pseudo-labelling}
\end{figure}

\noindent
\noindent Label propagation of satellite imagery is a technique employed to assign labels to unlabeled pixels or regions within satellite images, utilizing the available labeled samples. The process involves disseminating the known labels to neighboring or similar pixels in the image, thereby expanding the labeled dataset. To carry out label propagation in this work, we analyze the spectral characteristics of both labeled and unlabeled pixels, taking into account the similarity or proximity between pixels using the Jeffries-Matusita Spectral Angle Mapper (JMSAM) technique [\citenum{shaik2022automatic}]. This technique yields a score map ranging from 0 to 1. Leveraging this similarity information, the algorithm propagates labels from labeled pixels to neighboring or similar unlabeled pixels, as illustrated in figure \ref{fig:Pseudo-labelling}. The propagation is typically executed iteratively on each pixel within a 1 km radius buffer circle, gradually refining the labels of unlabeled pixels based on their neighboring pixels' labels. This iterative process was performed on all FIA plots, pseudo-labeling the pixels with a JMSAM score exceeding 0.99. In summary, the utilization of label propagation with satellite imagery and FIA plots increased the dataset size by ~3.5 times.

\FloatBarrier

The bar graph in figure \ref{fig:FIAFuelModels} shows the number of samples available for each fuel model in the dataset before label propagation. It can be observed from the figure that this dataset is imbalanced, which is known to cause difficulties in modeling with a significant disparity in the number of samples across different fuel models. Using an imbalanced dataset can have several implications for predictive modeling. First, it can introduce a bias towards the majority class, as the model may be more inclined to predict the dominant fuel model due to its higher representation. Second, the model's performance on the minority classes may be severely compromised, leading to lower accuracy and recall for those classes. Therefore, it is crucial to address the class imbalance issue to ensure more reliable and balanced predictions.

\begin{figure}[htbp]
  \centering
  \includegraphics[width=15cm, height=12.5cm]{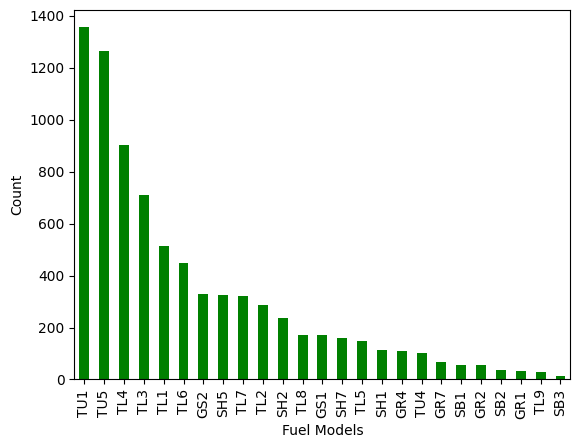} 
  \caption{Distribution of samples across different fuel models in the dataset after label propagation}
  \label{fig:imbalanced}
\end{figure}
\FloatBarrier
   
\subsection{Synthetic Data Generation}
\noindent We attempted to balance the dataset by augmenting it with synthetic data. Based on the existing literature [\citenum{endres2022synthetic}, \citenum{pathare2023comparison}, \citenum{GANSurvey}], we selected five methods for comparative analysis with our dataset---namely, Tabular Variational Autoencoders (TVAE), Conditional Tabular General Adversarial Networks (CTGAN), Synthetic Minority Oversampling Technique (SMOTE), Copula General Adversarial Networks (CGAN), and Gaussian Copula Synthesizer (GCS). Prior research suggested that these methods had demonstrable accuracy in other application fields [\citenum{GANSurvey}].

We trained the above-mentioned models using the Synthetic Data Vault library [\citenum{patki2016synthetic}, \citenum{lemaitre2017imbalanced}, \citenum{montanez2018sdv}] in Python and assessed them using various evaluation metrics, including overall quality score, column shapes, column pair trends, pairwise correlation distance, and proximity level. The definitions of the former three synthetic data evaluation metrics can be found in \citenum{endres2022synthetic}. These metrics evaluate the affinity of synthetic data to real data [\citenum{zhang2022sequential}] and enable a quantitative assessment of the quality of the generated/synthetic data. Values close to 0 indicate poor data quality, while values close to 1 indicate good data quality. 

Among these metrics, pairwise correlation distance concept is to ascertain whether the relationships among variables in the real data are maintained in the generated synthetic data. In order to do so, pearson correlation coefficients for real data and synthetic data were computed and stored in matrices $\textrm{df}_\textrm{real.corr}$ and $\textrm{df}_\textrm{synth.corr}$ respectively. Then the pairwise correlation distance is calculated as element-wise difference between two stored matrices as follows:

\begin{center}
\begin{equation}
\textrm{diff}_\textrm{corr} = \textrm{df}_\textrm{real.corr} - \textrm{df}_\textrm{synth.corr} 
\end{equation}
\end{center}

Then, heatmaps are generated utilizing the correlation distances of real and synthetic data to visualize and understand the correlation structure. When the correlation between two items is zero, it indicates that they are equivalent to each other, and vice versa [\citenum{endres2022synthetic}]. This analysis was carried to assess the fidelity of the correlation structure in synthetic data generated using different techniques.

To ensure a high-quality generated dataset, the value of 'diff' should be close to zero, indicating high proximity. Conversely, low-quality datasets will have proximity values significantly different from zero. Proximity, in this context, refers to a measure of similarity or dissimilarity between data points. The proximity value mentioned in this case refers to the average of pairwise correlation distances between real data and synthetic data [\citenum{bock2005proximity}].

\subsection{Importance of inclusive features}
\noindent A feature importance study can be carried out based on permutation importance. This involves permuting the column values of a single predictor feature and then passing all test samples back through the Random Forest to recompute the accuracy. The importance of that feature is determined by the difference between the baseline accuracy and the decrease in overall accuracy caused by permuting the column. While the permutation mechanism is considerably more computationally expensive than the mean decrease in impurity mechanism, the results are more reliable. The feature importance table \ref{tab:feature_imp} provides insightful information regarding the significance of various features. This table encompasses essential metrics, including importance score, standard deviation, p-value, p99-high, and p99-low. 

The importance score of a feature signifies the decrease in model performance based on perturbed data, where the values of this specific feature have been randomly shuffled across rows [\citenum{song2022dynamic}, \citenum{agdoc}]. For instance, a feature score of 0.01 indicates a predictive performance drop of 0.01 when the feature was randomly rearranged. The higher the score, the more critical the feature is for the model's performance. Conversely, a negative score implies that the feature is potentially detrimental to the final model, and removing it could enhance predictive performance. The standard deviation reflects the variability in a feature's importance across different model runs, where a low standard deviation indicates a consistent impact on predictions. The p-value helps determine the statistical significance of a feature's importance. A low p-value suggests that the feature significantly influences the target variable. For example, a p-value of 0.01 indicates a 1\% chance of the feature being useless or harmful and a 99\% chance of it being useful. Additionally, the p99-high and p99-low values specify the upper and lower bounds, respectively, for a feature's importance at the 99-th percentile. This range provides an estimate of the feature's likely impact. Overall, the feature importance table obtained is a valuable resource for comprehending the role and significance of different features within the predictive model.

\begin{table}[htbp]
    \centering
    \caption{Importance score of inclusive features in terms of overall accuracy}
    \label{tab:feature_imp}
    \begin{tabular}{p{0.5cm} p{2cm} p{2cm} p{2.0cm} p{2.4cm} p{2cm} p{2cm}}
         \toprule
        \textbf{No.} & \textbf{Variance} & \textbf{Importance} & \textbf{StdDev} & \textbf{p-value} & \textbf{p99\_high} & \textbf{p99\_low}\\
        \midrule
        1 & Elevation & 0.190246 & 0.012607	& 2.30072e-06 & 0.216205 & 0.164288 \\
        2 & NDVI\_3 & 0.111513 & 0.003718 & 1.48138e-07 & 0.119170 & 0.103857 \\
        3 & NDVI\_1 & 0.087381 & 0.001928 & 2.84011e-08 & 0.091350 & 0.083412 \\
        4 & NDVI\_4 & 0.080040 & 0.006472 & 5.08484e-06 & 0.093366 & 0.066715 \\
        5 & NDVI\_8 & 0.079638 & 0.005422 & 2.56236e-06 & 0.090802 & 0.068474 \\
        6 & NDVI\_5 & 0.074912 & 0.003695 & 7.07647e-07 & 0.082519 & 0.067305 \\
        7 & NDVI\_2 & 0.065058 & 0.003512 & 1.01531e-06 & 0.072289 & 0.057826 \\
        8 & NDVI\_6 & 0.060935 & 0.005431 & 7.49421e-06 & 0.072118 & 0.049752 \\
        9 & NDVI\_7 & 0.058019 & 0.002835 & 6.82073e-07 & 0.063857 & 0.052181 \\
        10 & Slope & 0.022624 & 0.005710 & 4.48223e-04 & 0.034382 & 0.010867 \\
        11 & PL\_ESPRIT & 0.022524 & 0.001301 & 1.33134e-06 & 0.025203 & 0.019844 \\
        12 & PL\_HV & 0.014882 & 0.002675 & 1.19981e-04 & 0.020389 & 0.009375 \\
        13 & S1\_VH & 0.012971 & 0.003544 & 6.07255e-04 & 0.020269 & 0.005673 \\
        14 & PL\_HH & 0.011262 & 0.002091 & 1.36325e-04 & 0.015568 & 0.006956 \\
        15 & S1\_SPAN & 0.004525 & 0.001629 & 1.71010e-03 & 0.007879 & 0.001170 \\
        16 & S1\_VV & 0.004324 & 0.002091 & 4.92799e-03 & 0.008630 & 0.000018 \\
        17 & PL\_NPDI & 0.003419 & 0.003419 & 9.24982e-03 & 0.007521 & -0.000683 \\
        18 & PL\_DIFF & 0.002614 & 0.002115 & 2.53267e-02 & 0.006970 & -0.001741 \\
        19 & PL\_Ratio & 0.002614 & 0.002145 & 2.63424e-02 & 0.007031 & -0.001802 \\
        20 & S1\_PRatio & 0.001508 & 0.000941 & 1.15252e-02 & 0.003445 & -0.000428 \\
        21 & S1\_D1 & 0.000101 & 0.001440 & 4.41731e-01 & 0.003065 & -0.002864 \\
        22 & S1\_VH\/VV & -0.000402 & 0.001252 & 7.43887e-01 & 0.002175 & -0.002980 \\
        23 & S1\_VV\/VH & -0.000804 & 0.001690 & 8.26413e-01 & 0.002675 & -0.004284 \\
        24 & S1\_RVI & -0.001106 & 0.001440 & 9.19529e-01 & 0.001858 & -0.004070 \\
        
        \bottomrule
    \end{tabular}
\end{table}
\FloatBarrier

Negative feature importance scores may be obtained, as we observe here for S1\_VHVV, S1\_VVVH, and S1\_RVI.  Although these scores are negative up to the fourth decimal digit, we do not dismiss them solely based on this observation and proceed to retrain the model by excluding these features. If the results remain unchanged---as they do in the present example---then, these features are kept within the model.

\subsection{Ensemble Model Training}
\begin{figure}[htbp]
  \centering
  \includegraphics[width=15cm, height=7.5cm]{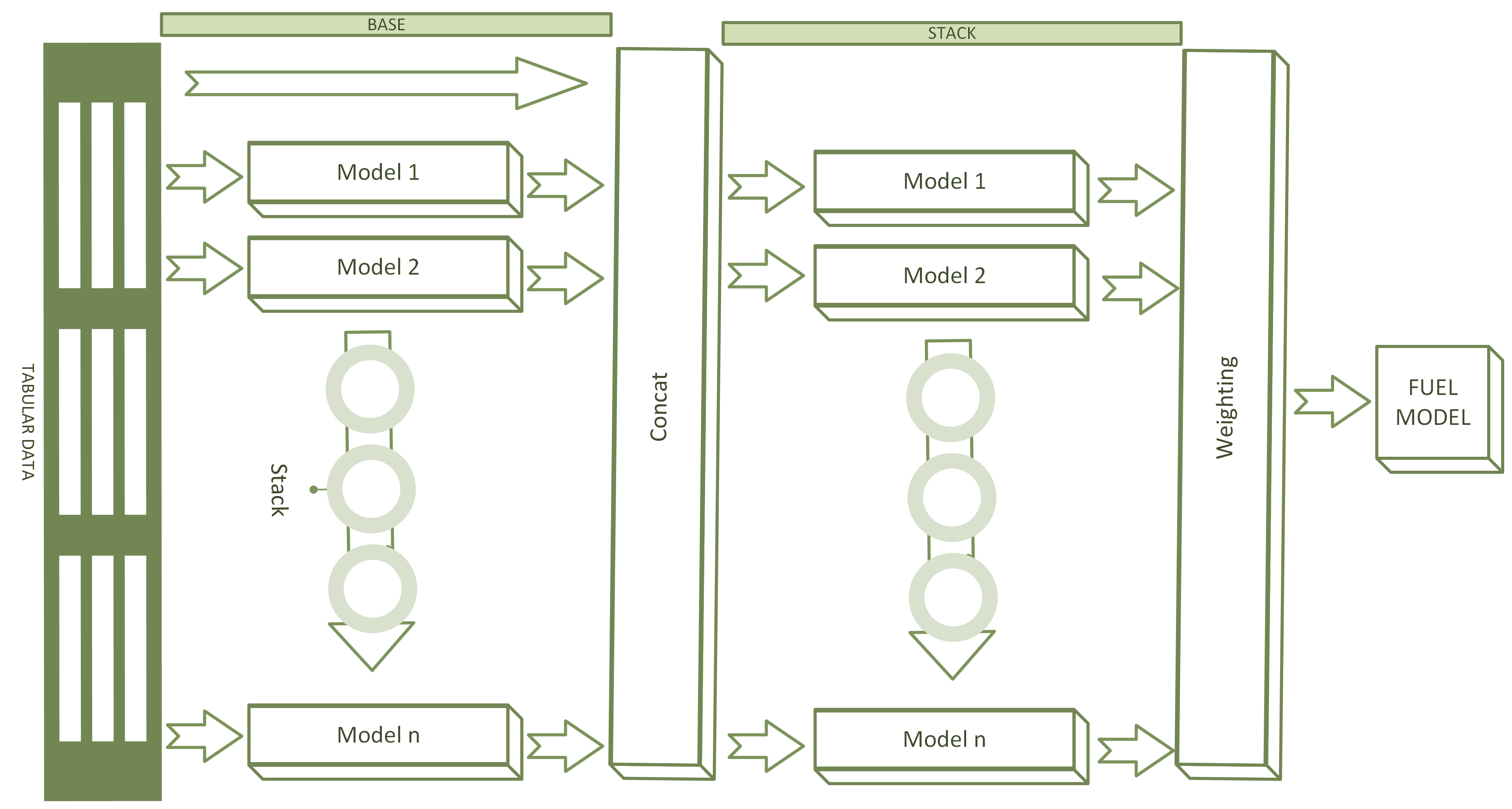} 
  \caption{Ensemble model with multi-layer stacking strategy, shown here using two stacking layers and 24 types of base learners.}
  \label{fig:autogluon}
\end{figure}

\noindent Our proposed ML architecture for fuel mapping is a heterogeneous ensemble model including neural networks, Light Gradient Boosting Method (GBM) boosted trees, CatBoost boosted trees, random forests, extremely randomized trees, and k-Nearest Neighbors. These models are combined with a stacking strategy wherein the base models learn in parallel, and a meta-model outputs the prediction based on the different base models' predictions. We implemented this procedure using the python-based ``gluon'' library [\citenum{autogluon, erickson2020autogluon}], which supports a multilayer stacking strategy that combines all models to enhance the performance of the model and excels in fine-tuning hyperparameters, surpassing the effectiveness of manual adjustments [\citenum{autogluon, agdoc}]. We train our ensemble model selecting on NVIDIA GeForce RTX 3070 GPU for 13,828 seconds (~4 hours). This meta-model utilizes 26 base models from the above-mentioned families (see, Table \ref{tab:validation}) for the purpose of ensembling and stacking as illustrated in \ref{fig:autogluon}.

\FloatBarrier

\begin{table}[htbp]
    \centering
    \caption{Validation and Accuracy Testing of the FuelVision Model}
    \label{tab:validation}
    \begin{tabular}{p{1cm} p{5.5cm} p{2.9cm} p{3.9cm}}
        \toprule
        \textbf{No.} & \textbf{Model} & \textbf{Test Acc} & \textbf{Validation Acc} \\
        \midrule
        1 & NeuralNetFastAI\_BAG\_L2 & 0.772750 & 0.761689 \\
        2 & LightGBMXT\_BAG\_L2 & 0.769734 & 0.776898 \\
        3 & WeightedEnsemble\_L3 & 0.769734 & +0.776898 \\
        4 & LightGBM\_BAG\_L2 & 0.767722 & 0.774007 \\
        5 & XGBoost\_BAG\_L2 & 0.763700 & 0.773002\\
        6 & CatBoost\_BAG\_L2 & 0.762695 & 0.769985\\
        7 & LightGBMLarge\_BAG\_L2 & 0.761187 & 0.769356\\
        8 & RandomForestGini\_BAG\_L2 & 0.753142 & 0.756662\\
        9 & ExtraTreesGini\_BAG\_L2 & 0.753142 & 0.746481\\
        10 & ExtraTreesEntr\_BAG\_L2 & 0.749120 & 0.738185\\
        11 & WeightedEnsemble\_L2 & 0.748115 & 0.749246\\
        12 & RandomForestEntr\_BAG\_L2 & 0.745601 & 0.748366 \\
        13 & NeuralNetTorch\_BAG\_L2 & 0.730015 & 0.710659 \\
        14 & LightGBMLarge\_BAG\_L1 & 0.725993 & 0.717320 \\
        15 & CatBoost\_BAG\_L1 & 0.720463 & 0.720714 \\
        16 & LightGBMXT\_BAG\_L1 & 0.720463 & 0.713298 \\
        17 & LightGBM\_BAG\_L1 & 0.715938 & 0.710533 \\
        18 & RandomForestGini\_BAG\_L1 & 0.714932 & 0.699723 \\
        19 & ExtraTreesGini\_BAG\_L1 & 0.713424 & 0.716817 \\
        20 & ExtraTreesEntr\_BAG\_L1 & 0.712921 & 0.713801 \\
        21 & NeuralNetTorch\_BAG\_L1 & 0.707391 & 0.691554 \\
        22 & XGBoost\_BAG\_L1 & 0.703871 & 0.689542 \\
        23 & RandomForestEntr\_BAG\_L1 & 0.701357 & 0.704500 \\
        24 & NeuralNetFastAI\_BAG\_L1 & 0.689794 & 0.649070 \\
        25 & KNeighborsDist\_BAG\_L1 & 0.359980 & 0.340372 \\
        26 & KNeighborsUnif\_BAG\_L1 & 0.317245 & 0.298140 \\
        
        \bottomrule
    \end{tabular}
\end{table}
\FloatBarrier

This advanced approach not only aids in minimizing overfitting but also enhances the overall system performance. The table reveals that the maximum difference between the testing and validation scores does not exceed 1.5\%, indicating minimal overfitting. By harnessing the collective intelligence of multiple models, this architecture effectively generalizes and accurately predicts outcomes on both the validation and testing datasets. The proposed ensemble and stacking methodology strengthens the model's robustness, ensuring dependable results even when faced with diverse data patterns and complexities.

\subsection{Non-Burnables Mapping: Post-Processing Step}
\noindent  The FIA field survey data only focuses on forest fuels and does not include non-burnables information. As a post-processing step, we map and label four types of non-burnable fuel models (NB1, NB2, NB8, and NB9) as NB. To map barren lands, snow, and water in the region of interest, we create NDVI and Normalized Difference Water Index (NDWI) [\citenum{wang2020urban}] images using the Google Earth Engine (GEE). Additionally, we generate a built-up index (BUI) [\citenum{rendana2023effects}], which is the difference between NDVI and Normalized Difference Build-up Index (NDBI), to map urban components. Utilizing the NDVI, NDWI, and BUI, we perform a post-processing step on the fuel map generated by the FuelVision model. Pixels with NDVI \textless 0, NDWI \textgreater 0.5, and BUI \textgreater 0.5 are labeled as NB. We then decide on these values by conducting experiments with a simple trial-and-error approach and visually checking the map. The equations for calculating NDWI and NDBI are provided in Eqs. (3 \& 4).

\begin{center}
\begin{equation}
NDWI = \frac{G-NIR}{G+NIR}
\end{equation}
\end{center}

\begin{center}
\begin{equation}
NDBI = \frac{MIR - NIR}{MIR+NIR}
\end{equation}
\end{center}

\begin{center}
\begin{equation}
BUI = NDVI - NDBI
\end{equation}
\end{center}

where G = Green, NIR = Near-infrared and MIR = Mid-infrared

\section{Results \& Discussions}
\noindent In this section, we carry out a comprehensive evaluation of the data augmentation techniques employed in our study, followed by a detailed model evaluation. We analyze the impact of pseudo-labeling and data augmentation techniques on the overall performance of our fire fuel mapping models. Additionally, fuel maps for select case studies are presented, showcasing the effectiveness of FuelVision in accurately classifying fuel types. Furthermore, we delve into the analysis of prediction uncertainties, leveraging prediction probabilities to assess the reliability and confidence of our model's predictions. The following subsections offer insights and discussions on each of these aspects and examine the robustness and effectiveness of the proposed methodology.

\subsection{Evaluation of Data Augmentation Techniques}
\noindent We assess all implemented data augmentation techniques using SD metrics, as described later in detail in \S4.2. The obtained values from the SD metrics are displayed in Table \ref{tab:syn_comp}, while the pairwise correlation distance heatmaps are presented in Figure \ref{fig:pairwise_corr}. We examined the metric values and compared them with non-SD evaluation metrics. The pairwise correlation distance heatmaps offer a visual representation of the relationships among features in the dataset, providing insights into both similarities and differences within the correlation patterns.

\begin{table}[htbp]
    \centering
    \caption{Comparative analysis of Synthetic Models}
    \label{tab:syn_comp}
    \begin{tabular}{p{1cm} p{2cm} p{2cm} p{2cm} p{2cm} p{2cm}}
        \toprule
        \textbf{No.} & \textbf{Models} & \textbf{Overall Quality Score} & \textbf{Column Shapes} & \textbf{Column Pair Trends} &\textbf{Proximity Level}\\
        \midrule
        1 & TVAE & 93.88 & 94.62 & 93.14 & -0.039 \\
        2 & CTGAN & 91.39 & 89.55 & 93.23 & -0.023\\
        3 & SMOTE & 90.81 & 88.75 & 92.88 & -0.026 \\
        4 & CGAN & 89.28 & 86.72 & 91.85 & -0.033 \\
        5 & GCS & 89.22 & 84.43 & 94.02 &  -0.021\\
        
        \bottomrule
    \end{tabular}
\end{table}
\FloatBarrier

Upon analyzing table \ref{tab:syn_comp}, it becomes evident that TVAE outperforms other models in terms of SD metrics, while GCS excels at proximity. However, upon visualizing the heatmaps of these two models in figure \ref{fig:pairwise_corr}, we discover that GCS is missing a feature, namely PL\_HV, which may be attributed to the data complexity---specifically, the non-Gaussian distribution of PL\_HV data.

\newgeometry{margin=1in}
\begin{landscape}

\begin{figure}[htbp]
  \centering
  \begin{adjustbox}{max width=\linewidth}
    \includegraphics[width=25cm, height=15cm]{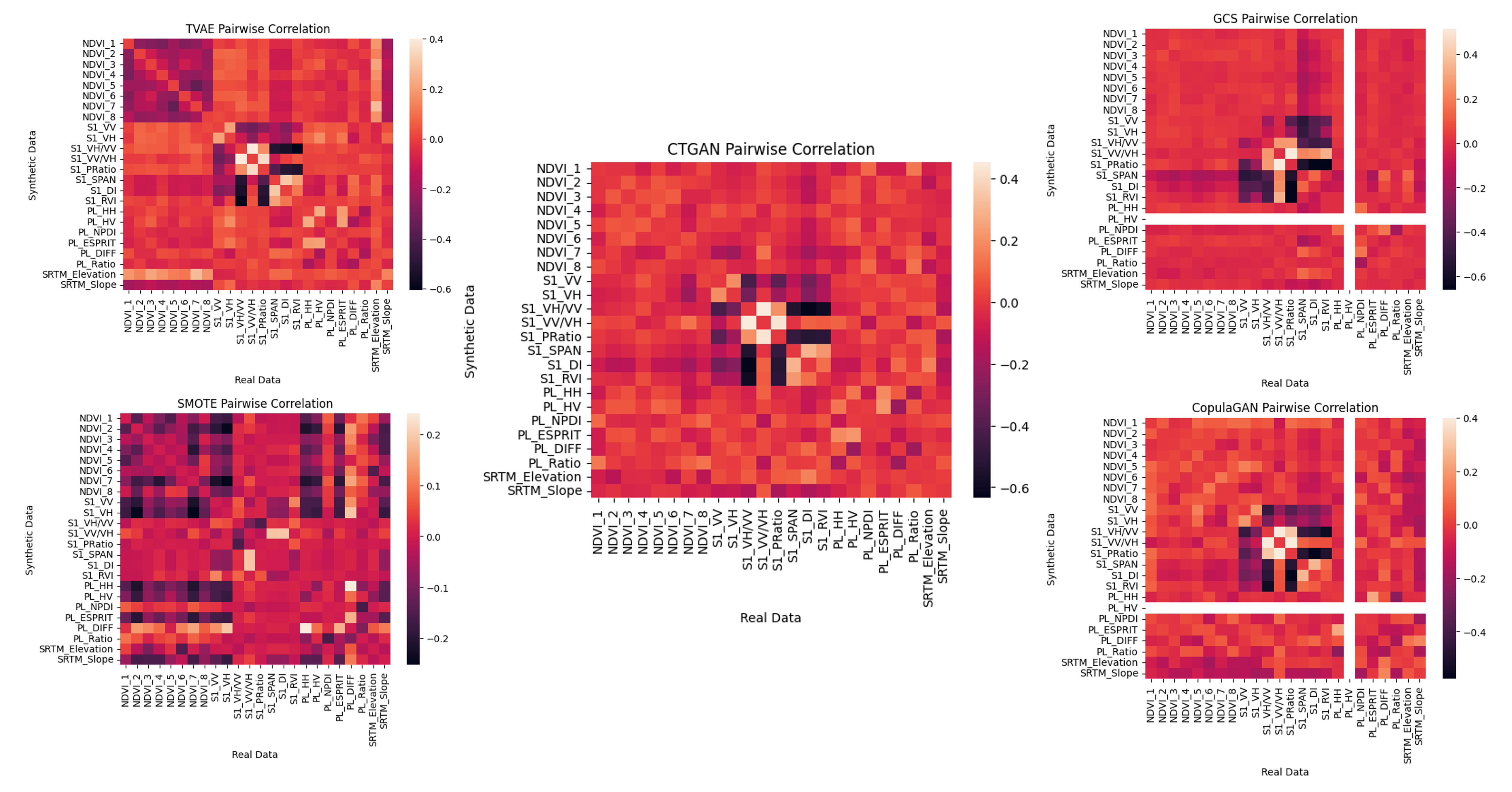} 
  \end{adjustbox}
  \caption{Pairwise Correlation distance for real and synthetic data}
  \label{fig:pairwise_corr}
\end{figure}
\FloatBarrier

\end{landscape}
\restoregeometry

Additionally, TVAE's heatmap appears less smooth in comparison. Consequently, we proceed to evaluate the CTGAN model, which yields the highest scores for both SD metrics (overall quality score of 91.39, column shapes value of 89.55, and column pair trends value of 93.23) and non-SD metrics (proximity level of -0.023). Furthermore, the heatmap generated by CTGAN exhibits noticeable improvements. As a result, CTGAN is selected as the optimal model for generating synthetic data to augment our training dataset. This approach allows us to increase the training dataset by ~10times to a total of 26,000 samples.

\subsection{Model Evaluation}
\begin{figure}[htbp]
  \centering
  \includegraphics[width=\textwidth]{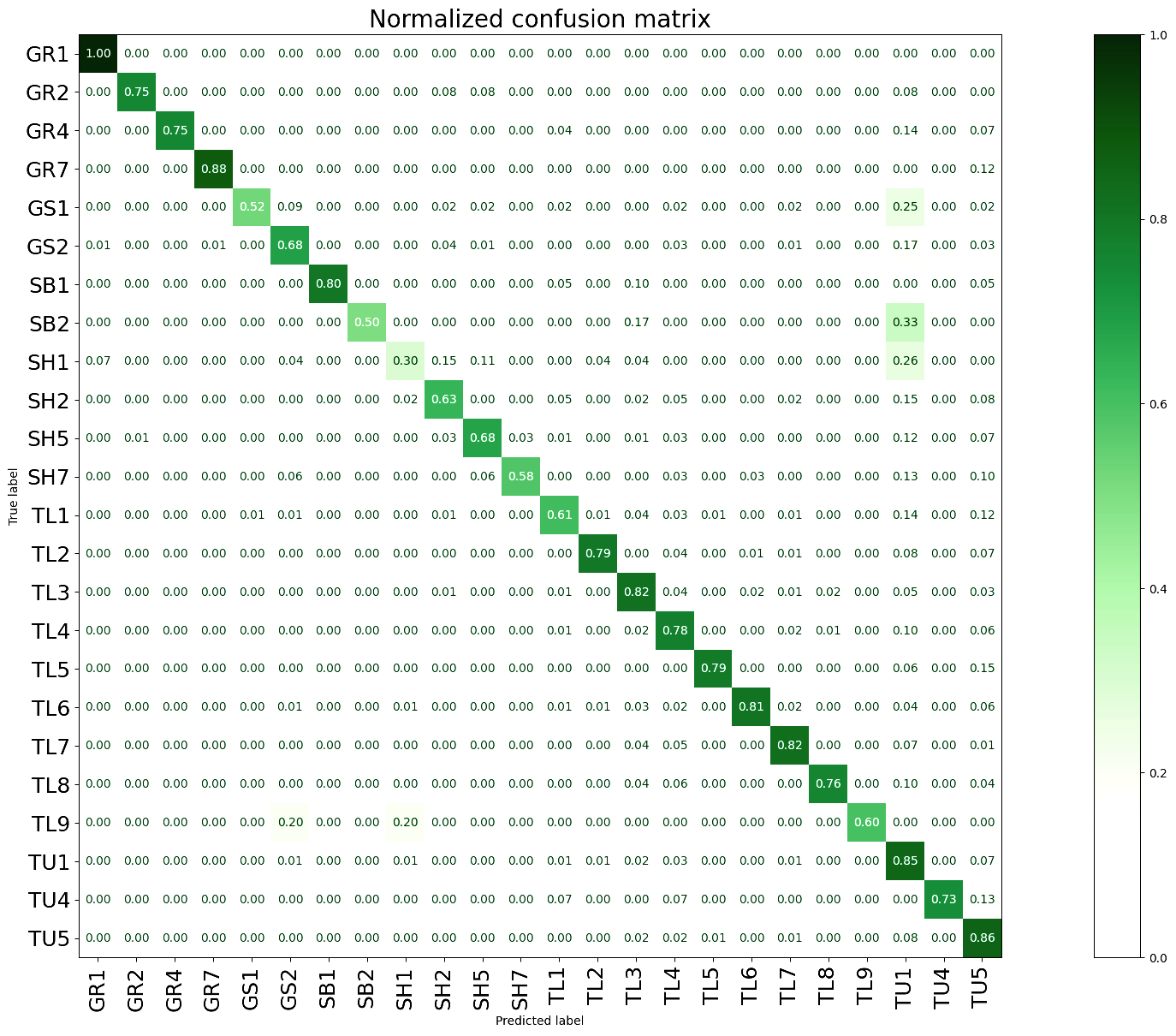} 
  \caption{Confusion matrix obtained with testing data by training ensemble model with FIA plots incorporated with pseudo-labels and synthetic data}
  \label{fig:conf_matrix_ovr}
\end{figure}

\noindent The FuelVision model was evaluated with a testing dataset of 1,989 samples with the performance metrics including precision, recall, and F1-score, as shown in Table \ref{tab:performancemetrics}. Therefore, these scores take false positives and false negatives into account together. F1-score is usually more beneficial than accuracy, especially when there is an uneven class distribution [\citenum{shaik2022automatic}]. Accuracy works best when false positives and false negatives have similar costs. 
The recall score, also known as sensitivity or true positive rate, measures the model's ability to correctly identify positive instances. In cases with different false positives and false negatives, it is better to consider precision and recall. The "Support" column in the table depicts the number of samples in the testing dataset. Due to insufficient ground truth data, an equal number of samples could not be considered. These performance metrics, along with the confusion matrix in figure \ref{fig:conf_matrix_ovr}, allowed for a comprehensive evaluation of FuelVision's accuracy, precision, recall, and overall effectiveness in predicting fuel models.

The f1-score for most of the fuel models is around 0.70, except for GS1 and SH1. Upon examining the confusion matrix in Figure \ref{tab:validation}, we see that GS1 and SH1 were misclassified as TU1. According to the definition, TU1 represents a fuel bed with a low load of grass and/or shrubs with litter, while GS1 and SH1 refer to shrubs that are 1 foot high, with the former having a low load of grass. There is a possibility that the low load of litter is insensitive to reflectance profiles or may appear as mixed pixels with a majority of shrubs and grasses. This could be the reason for misclassifying the GS1 and SH1 fuel models as TU1. Considering the presented performance metrics and confusion matrix, an overall accuracy of 0.77 is observed.

\begin{table}[htbp]
    \centering
    \caption{Performance Metrics on Test Dataset}
    \label{tab:performancemetrics}
    \begin{tabular}{p{1cm} p{2.5cm} p{2cm} p{2cm} p{2cm} p{2cm}}
        \toprule
        \textbf{No.} & \textbf{Fuel Model} & \textbf{Precision} & \textbf{Recall} & \textbf{f1-Score} &\textbf{Support}\\
        \midrule
        1 & GR1 & 0.50 & 1.0 & 0.67 & 4\\
        2 & GR2 & 0.90 & 0.75 & 0.82 & 12 \\
        3 & GR4 & 1.0 & 0.75 & 0.86 & 28 \\
        4 & GR7 & 0.78 & 0.88 & 0.82 & 8 \\
        5 & GS1 & 0.96 & 0.52 & 0.68 & 44 \\
        6 & GS2 & 0.79 & 0.68 & 0.73 & 76\\
        7 & SB1 & 1.0 & 0.80 & 0.89 & 20\\
        8 & SB2 & 1.0 & 0.50 & 0.67 & 6\\
        9 & SH1 & 0.62 & 0.30 &  0.40 & 27\\
        10 & SH2 & 0.75 & 0.63 & 0.68 & 65 \\
        11 & SH5 & 0.85 & 0.68 & 0.75 & 68\\
        12 & SH7 & 0.90 & 0.58 & 0.71 & 31\\
        13 & TL1 & 0.84 & 0.61 & 0.71 & 141 \\
        14 & TL2 & 0.92 & 0.79 & 0.85 & 92\\
        15 & TL3 & 0.80 & 0.82 & 0.81 & 173 \\
        16 & TL4 & 0.78 & 0.78 & 0.78 & 215\\
        17 & TL5 & 0.90 & 0.79 & 0.84 & 33\\
        18 & TL6 & 0.93 & 0.81 & 0.86 & 113\\
        19 & TL7 & 0.77 & 0.82 & 0.80 & 80\\
        20 & TL8 & 0.83 & 0.76 & 0.79 & 50\\
        21 & TL9 & 0.60 & 0.60 & 0.60 & 5\\
        22 & TU1 & 0.65 & 0.85 & 0.74 &358\\
        23 & TU4 & 1.00 & 0.73 & 0.85 & 15\\
        24 & TU5 & 0.73 & 0.86 & 0.79 &325\\
           & accuracy &  &  & 0.77 & 1989\\
           & macro avg & 0.82 & 0.72 & 0.75 & 1989\\
           & weighted avg & 0.79 & 0.77 & 0.77 & 1989\\
        
        \bottomrule
    \end{tabular}
\end{table}
\FloatBarrier

\FloatBarrier

\subsection{Impact of Pseudo-labelling \& Data Augmentation on Accuracy }
\begin{figure}[htbp]
  \centering
  \includegraphics[width=\textwidth]{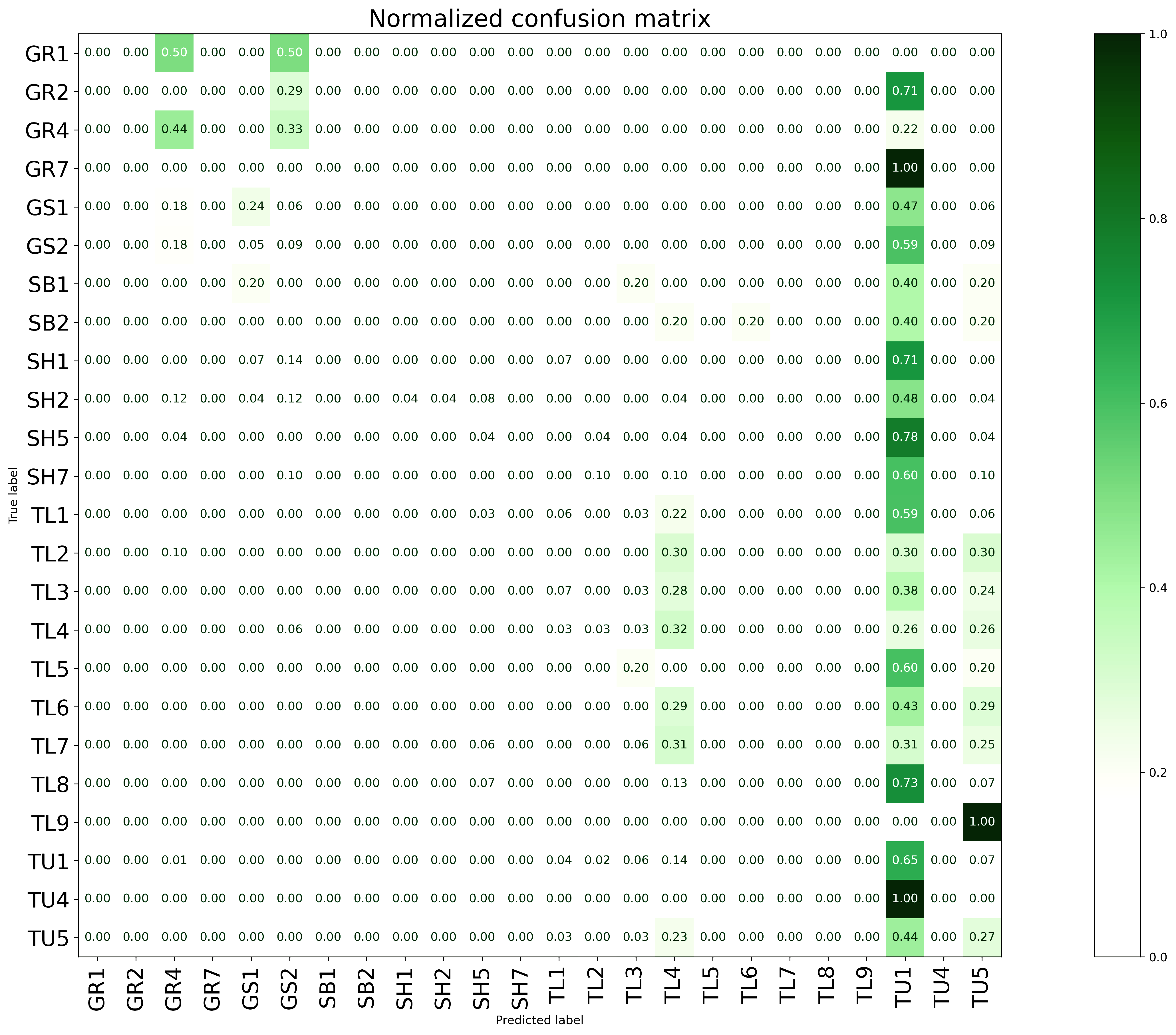} 
  \caption{Confusion matrix obtained with testing data by training ensemble model with exact FIA plots}
  \label{fig:conf_mat_exactlabels}
\end{figure}

\begin{figure}[htbp]
  \centering
  \includegraphics[width=\textwidth]{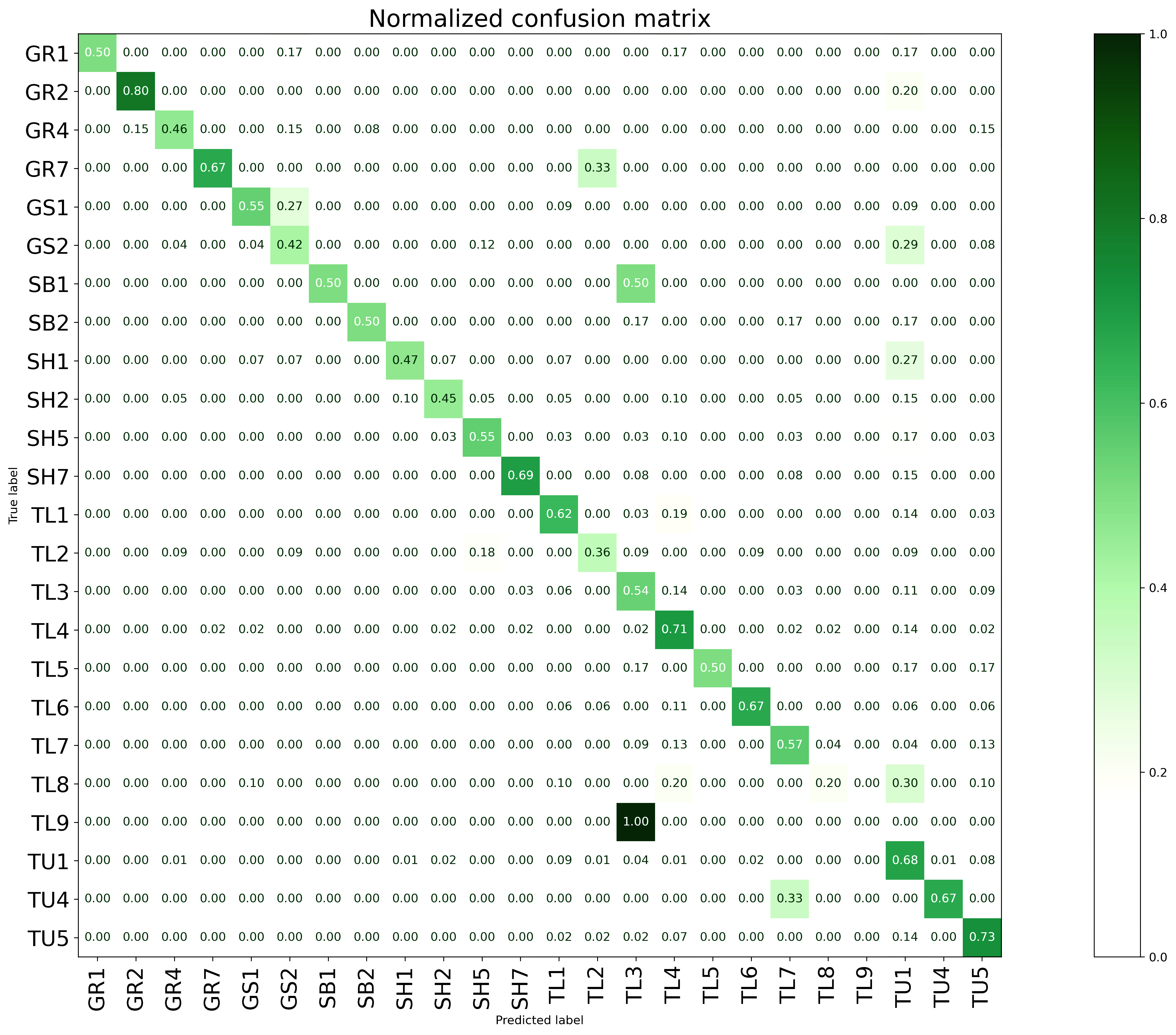} 
  \caption{Confusion matrix obtained with testing data by training ensemble model with FIA plots incorporated with pseudo-labels}
  \label{fig:conf_mat_pseudolabels}
\end{figure}

\noindent
The application of pseudo-labeling using the JM-SAM method and augmenting synthetic data generating using the CTGAN method enhances the accuracy of our FuelVision model, as shown in figures \ref{fig:conf_mat_exactlabels}, \ref{fig:conf_mat_pseudolabels} and \ref{fig:conf_matrix_ovr}. The confusion matrices shown in \ref{fig:conf_matrix_ovr} indicate remarkable improvements in the classification accuracy of each fuel model. Figure \ref{fig:conf_mat_exactlabels} depicts the evaluation results of the ensemble model trained with raw data, and it can be seen that there is no diagonal; instead, most of the fuel models are classified as TU1 as it has the higher number of plots, and can be correlated with FIA plots. However, after implementing pseudo-labeling and increasing the size of the training dataset, the diagonal accuracies experience a substantial boost (~20\%).

\FloatBarrier

\FloatBarrier

This increase is further amplified (~20\%) by incorporating data augmentation techniques, resulting in an even higher overall accuracy of 0.77 with strong diagonal accuracies. The initial training dataset exhibits a significant class imbalance, with TU1 being the majority class. Consequently, the trained model displays a bias towards classifying instances as TU1, resulting in an inflated number of misclassifications. However, after incorporating the pseudo-labels and augmenting the dataset with synthetic data, we observe a substantial reduction in this bias. The model's ability to generalize improves remarkably, as evidenced by a more balanced distribution of classifications across different classes. These advancements highlight the contributions of pseudo-labeling and augmentation with synthetic data, underscoring their effectiveness in refining our fuel mapping model.

\subsection{Fuel Map Generation for Case Studies}
\noindent The severity of wildfires in the Western United States has been steadily increasing since the mid-1980s, posing greater risks to human lives, properties, carbon storage, biodiversity, and other vital ecosystem services. In the past decade, the expansion of wildfire incidents has accelerated, culminating in an unprecedented fire season in 2020. This particular year witnessed over 2.5 million acres burning in the Western United States, with California accounting for 38\% of that total [\citenum{mcwethy2019rethinking}]. Since 2018, California has endured a series of fire seasons that broke previous records in burned area and losses. More than 27,000 homes and commercial structures have been destroyed, and the costs associated with fire suppression have soared [\citenum{palinkas2020california}]. In this study, we considered two representative California wildfire events for the development of fuel maps, which will be described below and will be incorporated into wildfire simulations in the near future.

The \textit{Dixie fire}, which began on July 13, 2021, holds the record as the largest documented wildfire to date, scorching an area of 374,000 hectares. The cost of suppressing this particular wildfire surpassed the \$500 million mark for the first time, reaching \$637 million in 2021 [\citenum{taylor2022severity}]. This fire's rapid spread and exceptionally intense behavior have piqued the interest of numerous fire scientists, prompting numerous studies on fire simulation. Fuel mapping plays a vital role in understanding and effectively managing wildfire risks. By accurately evaluating fuel conditions, fire behavior can be better predicted, strategies for fire suppression can be optimized, and overall fire management endeavors can be improved. The region where the Dixie fire occurred is considered a significant case for the development of fuel maps.

The second test case, known as the \textit{Caldor fire}, continued on for 67 days, burning through approximately 221,835 acres of land (including 9,885 acres within the Lake Tahoe Basin), destroying 1,003 structures and necessitating the evacuation of over 50,000 residents [\citenum{sion2023assessment, east2022analysis}]. The confluence of regional drought, intense heat, and powerful winds resulted in highly active fire behavior, giving rise to additional wildfires known as spot fires. These spot fires exacerbated the fire's expansion and posed additional challenges for firefighters in their containment efforts. [\citenum{wadhwani2022review}]

\begin{figure}[htbp]
  \centering
  \includegraphics[width=\textwidth]{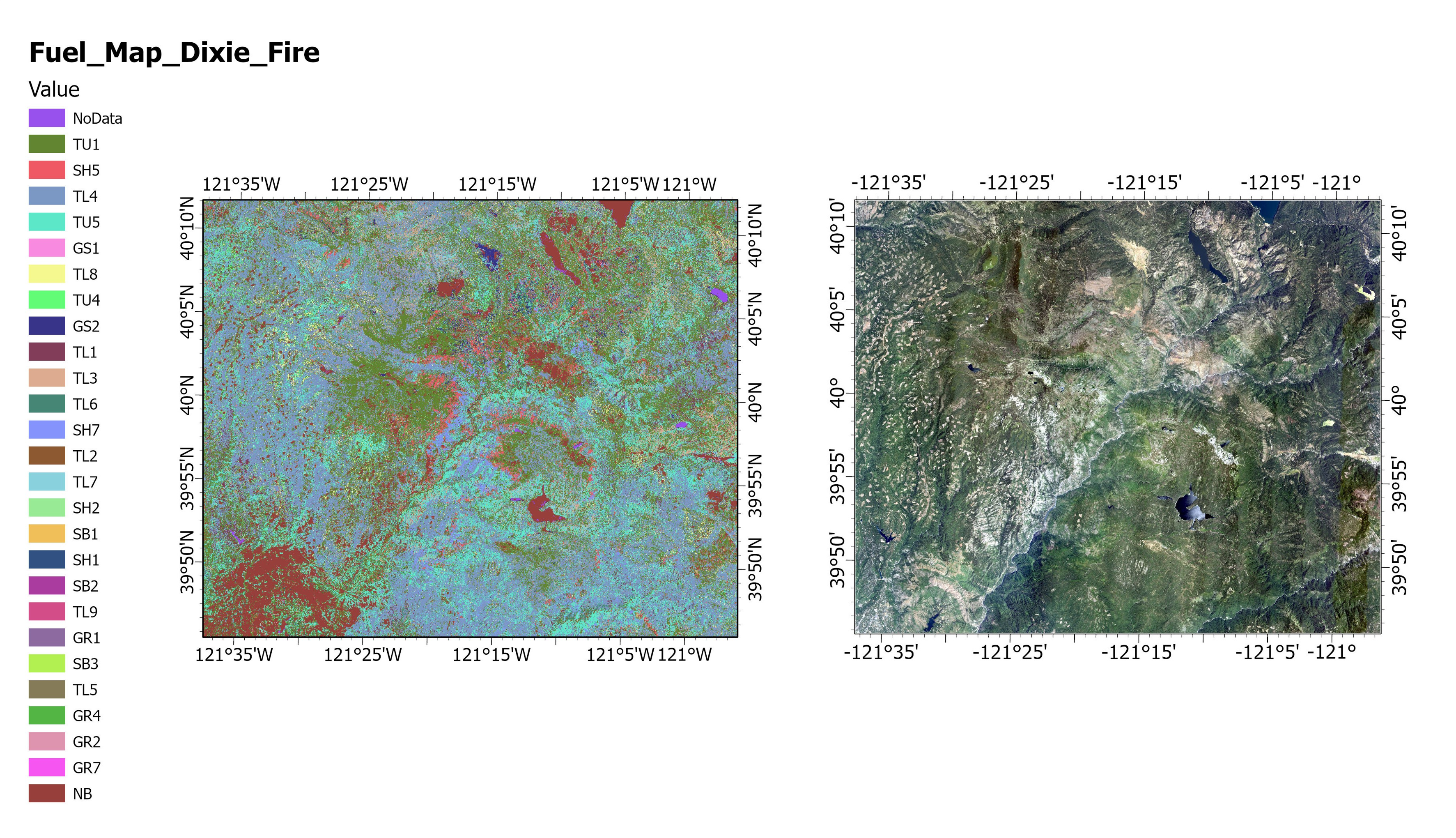} 
  \caption{(a) Fuel Map and (b) NAIP Image for the region of Dixie Fire}
  \label{fig:dixie_fire}
\end{figure}
\FloatBarrier

The process employed to generate these fuel maps for the Dixie and Caldor fires' regions of interest utilizing FulVision is outlined as follows:\newline
1. The input data for the region of interest is collected as outlined in Table \ref{tab:input_data} using Google Earth Engine.\newline
2. The remote sensing data is pre-processed, and spectral indices, as specified in Table \ref{tab:spectral_indices}, are calculated using Python.\newline
3. The satellite imagery and spectral indices are converted into a tabular format and saved as a CSV file, serving as the training dataset.\newline
4. Employing the generated CSV file and FuelVision, fuel models are predicted for each pixel parallely.\newline
5. A post-processing algorithm is applied to map the non-burnables as described in \S3.5 and transforms the CSV file into a GTiff file.\newline

The proposed FuelVision model successfully maps all the fuel models in the Dixie and Caldor fire areas, which are presented in Figures 11(a) and 12(a), respectively. Additionally, Figures 11(b) and 12(b) display the high-resolution imagery from NAIP for visual comparison. These high-resolution images allow us to compare the predicted fuel maps, specifically in terms of vegetation cover and vegetation patterns. These case studies can serve as a reference guide for creating fuel maps for any region of interest in California.

\begin{figure}[htbp]
  \centering
  \includegraphics[width=\textwidth]{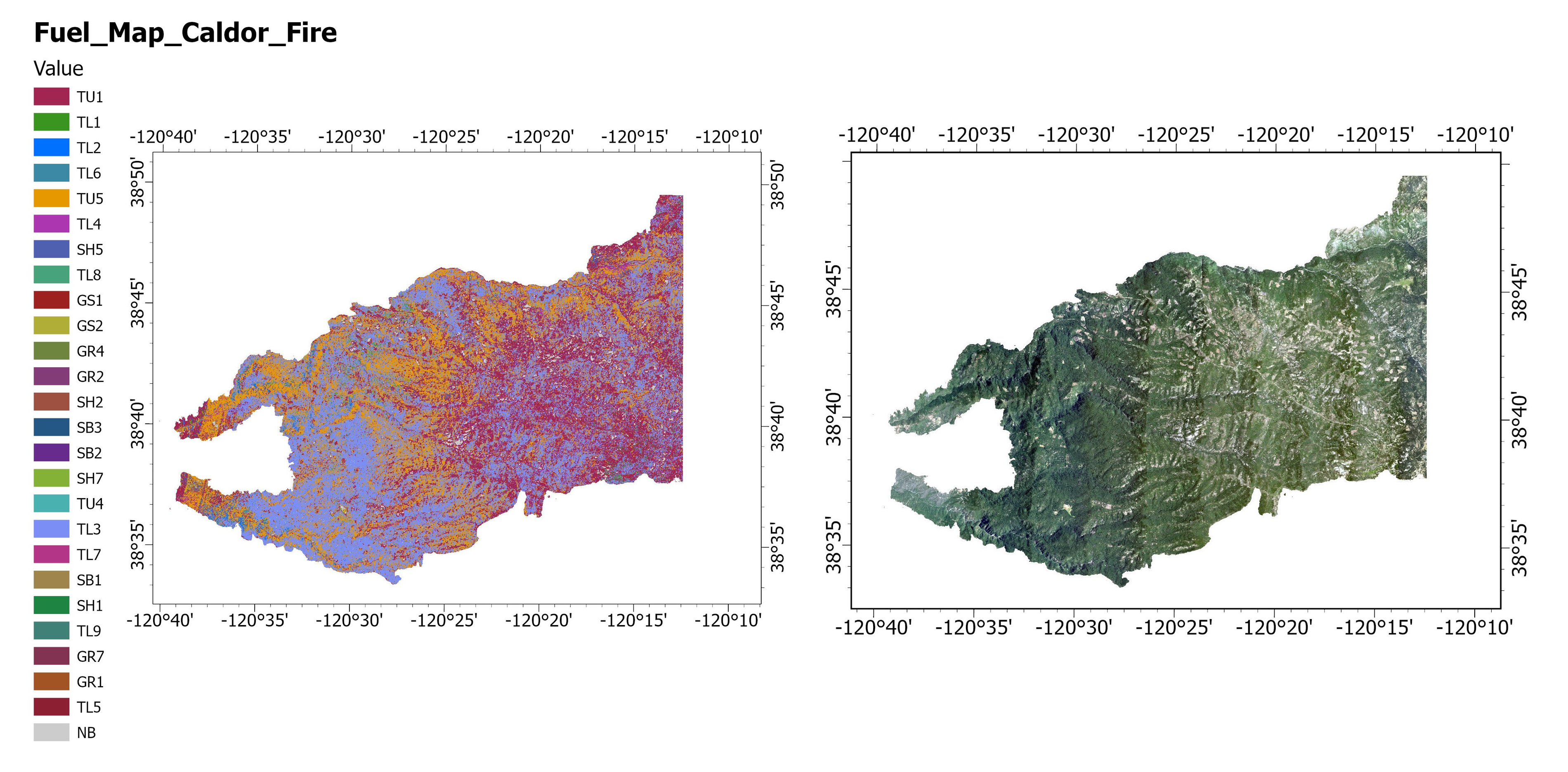} 
  \caption{(a) Fuel Map and (b) NAIP Image for the region of Caldor Fire}
  \label{fig:caldor_fire}
\end{figure}
\FloatBarrier

\subsection{Model Uncertainty Analysis}
\noindent Model uncertainty plays a crucial role in evaluating the reliability and confidence of predictions made by machine learning models. Analyzing the prediction probabilities provided by the model allows us to assess the level of uncertainty associated with each prediction. Our study generated prediction probabilities for the Dixie fire's area, as illustrated in Figure \ref{fig:dixie_fire}. These probabilities range from 0.01 to 0.99. They represent the model's estimation of the likelihood of a given instance belonging to a particular class. Higher prediction probabilities indicate a greater level of certainty, suggesting that the model has more confidence in its predictions. Conversely, lower prediction probabilities imply a higher degree of uncertainty, indicating that the model may be less confident or conflicted in its classification decision.

\begin{figure}[htbp]
  \centering
  \includegraphics[width=16cm]{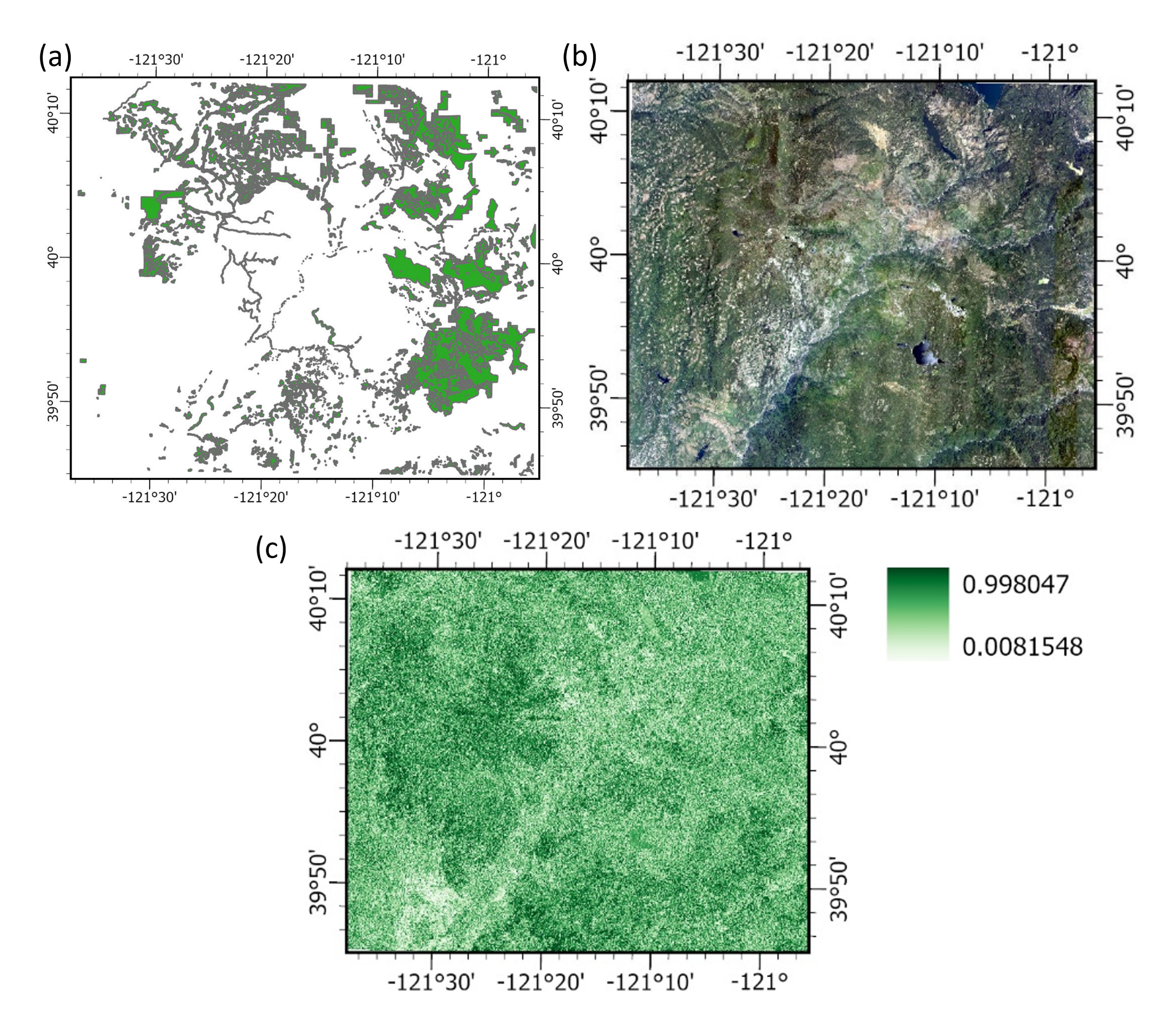} 
  \caption{Analysis of model uncertainty using (a) timber harvest map for Dixie fires region, (b) NAIP high-resolution imagery for the same region and (c) Dixie fire's fuel prediction probability map.}
  \label{fig:uncertain}
\end{figure}
\FloatBarrier

In our case, we start the analyses by comparing non-burnables (NBs) since the model was not trained on all NBs, as mentioned in the previous section, but they were mapped in the post-processing steps. Consequently, the model predicted these areas with lower probabilities. Referring to the NAIP image in figure \ref{fig:uncertain}, we observe that the southwest region of interest contains non-burnables such as rocks, barren lands, and roadways passing through the center. We notice a similar pattern in the probabilities map, where these areas showed lower probabilities since they were mapped as fuels in the fuel map.

Next, we compare the vegetated parts, which are primarily located on the south and west sides. Correspondingly, the probabilities in these regions are high ($> 0.9$). Towards the northeast, we observe slightly lower probabilities ranging from 0.7 to 0.9. Further analysis reveals that these areas roughly corresponded to the timber harvest locations indicated on the timber harvest map, as shown in Figure \ref{fig:uncertain}. This analysis of the prediction probabilities provided valuable insights into model uncertainty, enabled quantified interpretations of the varying degrees of confidence in downstream predictions, accounted for post-processing mapping effects, and revealed correlations between probability patterns and landscape features.

\section{Conclusions}
\noindent
Most existing studies on wildfire fuel mapping focus on developing models that are trained and applicable to small areas of interest. In contrast, this paper presents a model for mapping wildfire fuels in real-time at any selected domain, including large scales such as regional or state levels. The proposed model leverages AutoML to create a predictive model that combines information from optical, SAR, and terrain data. Specifically, we opted to devise an ensemble model with a multilayer stacking strategy that is known for its ability to classify multiple classes. This model utilizes a multi-model stacked ensemble approach, enhancing model performance while also providing a measure of uncertainty for the predicted fuels.

To evaluate the proposed system, we applied it to a dataset labelled with FIA plots and then improved by pseudo-labeling and augmenting synthetic data. The fuel labels for the state of California were based on the Scott and Burgan 40 fuel models. Synthetic data was generated using the CTGAN technique, and we conducted a comparative analysis using TVAE, GCS, CGAN, and SMOTE to select the most suitable method. Additionally, we performed a feature importance analysis to assess the impact of each feature on accuracy. The final results demonstrated the feasibility of the proposed approach, achieving an overall fuel classification accuracy of 77\% on an independent test set. Furthermore, analyzing the properties of the system revealed that fusing different modalities of data improves identification accuracy compared to using each data source individually.

To demonstrate the effectiveness of the proposed approach, we considered two cases: the Dixie and Caldor fires. Fuel maps were developed for these cases, and uncertainty analysis was conducted using the Dixie fire fuel map. The probabilities obtained from the model were found to correspond well with the landscape features. This procedure indicates that the proposed approach can be employed for large-scale real-time wildfire fuel mapping.


\section{Acknowledgements}
\noindent This work was supported through the NSF grant CMMI-1953333. The authors Shaik and Taciroglu also acknowledge the additional support provided through UCLA’s Amazon Science Hub Program and Edison International. The authors would like to thank Hans Anderson and John Chase from USDA Forest Services for providing the FIA data and assisting in its interpretation. Also, the authors gratefully acknowledge the support of the NVIDIA Corporation for this research. The opinions and perspectives expressed in this study are those of the authors and do not necessarily reflect the views of the sponsor. 


\bibliographystyle{elsarticle-num}
\bibliography{cas-refs} 





\end{document}